\documentclass[aps,prb,twocolumn,showpacs,superscriptaddress,groupedaddress]{revtex4} 

\usepackage{graphicx,color,float}  
\usepackage[section]{placeins}

\newcommand{\zco}{ZnCr$_2$O$_4$}
\newcommand{\zcuco}{Zn$_{0.9}$Cu$_{0.1}$Cr$_2$O$_4$}
\newcommand{\zcucuco}{Zn$_{0.8}$Cu$_{0.2}$Cr$_2$O$_4$}
\newcommand{\zcoco}{Zn$_{0.9}$Co$_{0.1}$Cr$_2$O$_4$}
\newcommand{\zcococo}{Zn$_{0.8}$Co$_{0.2}$Cr$_2$O$_4$}
\newcommand{\mco}{MgCr$_2$O$_4$}
\newcommand{\mcuco}{Mg$_{0.9}$Cu$_{0.1}$Cr$_2$O$_4$}
\newcommand{\mcucuco}{Mg$_{0.8}$Cu$_{0.2}$Cr$_2$O$_4$}
\newcommand{\aco}{$A$Cr$_2$O$_4$}
\newcommand{\cco}{CoCr$_2$O$_4$}
\newcommand{\cuco}{CuCr$_2$O$_4$}

\begin{document}

\title{Structural ground states of ($A,A'$)Cr$_2$O$_4$($A$=Mg, Zn; $A^{\prime}$ = Co, Cu) 
spinel solid solutions: Spin-Jahn-Teller and Jahn-Teller effects}

\author{Moureen C. Kemei}\email{kemei@mrl.ucsb.edu}
\author{Stephanie L. Moffitt}
\author{Lucy E. Darago}
\author{Ram Seshadri}\email{seshadri@mrl.ucsb.edu}
\affiliation{Materials Department and Materials Research Laboratory\\
University of California, Santa Barbara, CA, 93106, USA}
\author{Matthew R. Suchomel}\email{suchomel@aps.anl.gov}
\affiliation{Advanced Photon Source, Argonne National Laboratory, Argonne IL, 60439, USA}
\author{Daniel P. Shoemaker}\email{dpshoema@illinois.edu}
\affiliation{Department of Materials Science and Engineering\\ 
University of Illinois at Urbana-Champaign, Urbana IL 61801, USA}
\author{Katharine Page}\email{kpage@lanl.gov}
\author{Joan Siewenie}\email{siewenie@lanl.gov}
\affiliation{Lujan Neutron Scattering Center, Los Alamos National Laboratory, Los Alamos, New Mexico  87545, USA}

\date{\today}  

\begin{abstract}
We examine the effect of small amounts of magnetic substituents in the
$A$ sites of the frustrated spinels MgCr$_2$O$_4$ and ZnCr$_2$O$_4$. Specifically
we look for the effects of spin and lattice disorder on structural changes
accompanying magnetic ordering in these compounds. 
Substitution of Co$^{2+}$ on the non-magnetic Zn$^{2+}$ site
in Zn$_{1-x}$Co$_{x}$Cr$_2$O$_4$ where 0\,$<$\,$x$\,$\leq$\,0.2 completely
suppresses the spin-Jahn-Teller distortion of ZnCr$_2$O$_4$ although these systems
remain frustrated, and magnetic ordering occurs at very low temperatures of
$T$\,$<$\,20\,K. On the other hand, the substitution of Jahn-Teller
active Cu$^{2+}$ for Mg$^{2+}$ and Zn$^{2+}$ in
Mg$_{1-x}$Cu$_{x}$Cr$_2$O$_4$ and Zn$_{1-x}$Cu$_{x}$Cr$_2$O$_4$ where
0\,$<$\,$x$\,$\leq$\,0.2  induce Jahn-Teller ordering at temperatures well
above the N\'eel temperatures of these solid solutions, and yet spin interactions remain
frustrated with long-range magnetic ordering occurring below 20\,K without any
further lattice distortion. The Jahn-Teller distorted solid solutions
Mg$_{1-x}$Cu$_{x}$Cr$_2$O$_4$ and Zn$_{1-x}$Cu$_{x}$Cr$_2$O$_4$ adopt the
orthorhombic $Fddd$ structure of ferrimagnetic \cuco. Total neutron scattering
studies of Zn$_{1-x}$Cu$_{x}$Cr$_2$O$_4$ suggest that there are local $A$O$_4$
distortions in these Cu$^{2+}$-containing solid solutions at room temperature 
and that these distortions become cooperative when average structure 
distortions occur. Magnetism evolves from compensated antiferromagnetism in 
\mco\, and \zco\, to uncompensated antiferromagnetism with substitution of 
magnetic cations on the non-magnetic cation
sites of these frustrated compounds. The sharp heat capacity anomalies associated
with the first-order spin-Jahn-Teller transitions of \mco\, and \zco\, become broad
in Mg$_{1-x}$Cu$_{x}$Cr$_2$O$_4$, Zn$_{1-x}$Co$_{x}$Cr$_2$O$_4$, and
Zn$_{1-x}$Cu$_{x}$Cr$_2$O$_4$ when $x$\,$>$\,0. We present a temperature-composition
phase diagram summarizing the structural ground states and magnetic properties of
the studied spinel solid solutions.

\end{abstract}

\maketitle

Triangular lattice topologies are at the center of complex ground states in
functional oxides as has been shown in the charge ordered triangular metallic
AgNiO$_2$ where charge ordering rather than a Jahn-Teller distortion lifts orbital
degeneracy\cite{pascut_2011} and in geometrically frustrated spin systems such as
ZnCr$_2$O$_4$ where magnetic ordering is accompanied by a lattice
distortion.\cite{lee_2002} The ground states of the canonical spin frustrated
systems \aco\, ($A$\,=\,Mg,\cite{ehrenberg_2002,martin_2008,rovers_2002,kemei_2013}
Zn,\cite{lee_2002,ramirez_1994,kagomiya_2002, kemei_2013}
Cd,\cite{rovers_2002,chung_2005,aguilar_2008} and Hg\cite{Ueda_2006}) have been
extensively explored. To understand the degenerate ground states of ACr$_2$O$_4$
spinels, the effect of spin disorder on the magnetic properties of these systems has
been investigated; spin disorder is introduced by substituting magnetic ions on the
non-magnetic $A$ sublattice of these materials.\cite{melot_2009, kemei_2012,
yan_2008} Similarly, the effect of low concentrations of magnetic vacancies on the
Cr sublattice of ZnCr$_{2(1-x)}$Ga$_{2x}$O$_4$ has been studied showing that the
freezing temperature of these systems for small $x$ is independent of the spin
vacancy concentration.\cite{laforge_2013} However, the effect of spin and lattice
disorder on the structural ground states of the canonical frustrated systems \mco\,
and \zco\, has so far not been studied.

Here, we study magnetic ordering and correlated or uncorrelated structual changes
in \mco\, and \zco\, when low concentrations of magnetic cations are substituted 
on the non-magnetic $A$ site.  \mco\, and \zco\, are ideal candidates for the present 
study as they are: (i) Strongly frustrated with expected ordering temperatures of about 400\,K yet
suppressed antiferromagnetic ordering occurs below 15\.K\cite{dutton_2011} (ii)
Their structural and magnetic ground states are strongly coupled with a lattice
distortion occurring concomitantly with antiferromagnetic
ordering.\cite{ehrenberg_2002,lee_2002,kemei_2013} (iii) Finally, Cr$^{3+}$ 3$d^3$ strongly
prefers the octahedral site where it has a non-degenerate electronic configuration
thus enabling compositional variation only on the tetrahedral $A$ site. The effect
of spin disorder on the structural ground states of \zco\, is investigated by
substituting magnetic Co$^{2+}$ with a tetrahedral ionic radius of 0.58\,\AA\, for
Zn$^{2+}$ which has an ionic radius of 0.6\,\AA\, in tetrahedral
coordination.\cite{shannon_1976} The similarity in ionic radii between Co$^{2+}$ and
Zn$^{2+}$ minimizes the effects of lattice distortion while allowing us to probe the
effect of dilute $A$ site spins on the structural ground states of \zco. 
Jahn-Teller active Cu$^{2+}$ with an ionic radii of 0.57\,\AA\, is introduced to the
$A$ sites of \mco\, and \zco\, to study the effect of both structural and spin
disorder on the structural ground states of these systems.\cite{shannon_1976}
Mg$^{2+}$ has a Shannon-Prewitt ionic radius of 0.57\,\AA\, in tetrahedral
coordination.\cite{shannon_1976} 

This study is enabled by variable-temperature high-resolution synchrotron X-ray
diffraction, which is a powerful tool for investigating the coupling of spin and
lattice degrees of freedom in magnetic oxides. For example, it has been used to show
that exchange striction drives further distortions to orthorhombic symmetry in the
already Jahn-Teller distorted tetragonal spinels NiCr$_2$O$_4$ and
CuCr$_2$O$_4$.\cite{suchomel_2012} Similarly, high-resolution synchrotron X-ray
diffraction revealed phase coexistence in the spin-Jahn-Teller phases of \mco\, and
\zco.\cite{kemei_2013} Barton $et\,al.$ have also shown a spin-driven rhombohedral
to monoclinic structural distortion in Co$_{10}$Ge$_3$O$_{16}$.\cite{barton_2013} 

We show that $\geq$\,10$\%$ Co$^{2+}$ ions on the Zn$^{2+}$ site of \zco\, suppress
the structural distortion that accompanies antiferromagnetic ordering in \zco. We
also find that concentrations $\geq$\,10$\%$ of Jahn-Teller active Cu$^{2+}$ on the
Mg$^{2+}$ site of \mco\, and on the Zn$^{2+}$ site of \zco\, induce average
structure distortions at temperatures above the magnetic ordering temperature.  The
Jahn-Teller average structure distortion in Mg$_{1-x}$Cu$_x$Cr$_2$O$_4$ and
Zn$_{1-x}$Cu$_x$Cr$_2$O$_4$ occurs at higher temperatures with increase in $x$.
Despite the lattice distortions in Mg$_{1-x}$Cu$_x$Cr$_2$O$_4$ and
Zn$_{1-x}$Cu$_x$Cr$_2$O$_4$ when $x$\,$\geq$\,0.1, magnetic interactions remain
frustrated and no further average structure distortions are observed at the N\'eel
temperature. The Jahn-Teller distorted systems Mg$_{1-x}$Cu$_x$Cr$_2$O$_4$ and
Zn$_{1-x}$Cu$_x$Cr$_2$O$_4$ when $x$\,$\geq$\,0.1, are orthorhombic in the space
group $Fddd$. In all the studied solid solutions, magnetism evolves from frustrated
antiferromagnetism to glassy uncompensated antiferromagnetism. 

\section{Methods}
Powder samples of Zn$_{1-x}$Co$_x$Cr$_2$O$_4$, Mg$_{1-x}$Cu$_x$Cr$_2$O$_4$, and
Zn$_{1-x}$Cu$_x$Cr$_2$O$_4$ were prepared using solid state preparation methods. The
samples Mg$_{1-x}$Cu$_x$Cr$_2$O$_4$ were prepared from stoichiometric solution
mixtures of the nitrates Mg(NO$_3$)$_2$.6H$_2$O, Cu(NO$_3$)$_2$.6H$_2$O, and
Cr(NO$_3$)$_3$.9H$_2$O. The nitrate precursor was calcined at temperatures between
700\,$^{\circ}$C and 1000\,$^{\circ}$C for 10 hours as reported by Shoemaker and
Seshadri.\cite{shoemaker_2010} Powders of Zn$_{1-x}$Cu$_x$Cr$_2$O$_4$ were prepared
from stoichiometric amounts of ZnO, CuO, and Cr$_2$O$_3$ that were ground, pressed
into pellets, and calcined at 800\,$^{\circ}$C for 12 hours. These samples were
reground, pressed into pellets, and annealed at 1000\,$^{\circ}$C for 48 hours
followed by further annealing at 800\,$^{\circ}$C for 12 hours. Stoichiometric
powders of Zn$_{1-x}$Co$_x$Cr$_2$O$_4$ were prepared from CoC$_2$O$_4$.2H$_2$O, ZnO,
and Cr$_2$O$_3$. These powders were mixed, pressed into pellets, and calcined at
800\,$^{\circ}$C for 12 hours. The samples were then reground, pressed into pellets,
and annealed at 1150\,$^{\circ}$C for 12 hours followed by further annealing at
800\,$^{\circ}$C for 24 hours. Samples were structurally characterized by
high-resolution ($\delta Q$/$Q\,\leq\,$2$\,\times$10$^{-4}$) variable-temperature
(6\,K\,$\leq$\,$T$\,$\leq$\,300\,K) synchrotron X-ray powder diffraction at beamline
11BM at the Advanced Photon Source, Argonne National Laboratory. Diffraction
patterns were fit to structural models using the Rietveld method as implemented in
the EXPGUI/GSAS software program.\cite{toby_expgui_2001,larson_2000} Crystal
structures are visualized using the program VESTA. \cite{momma_vesta_2008} Magnetic
properties were measured using the Quantum Design MPMS 5XL superconducting quantum
interference device (SQUID). Heat capacity measurements were performed using a
Quantum Design Physical Properties Measurement System. Time-of-flight neutron
scattering data was collected on the NPDF instrument at Los Alamos National
Laboratory. The neutron pair distribution function (NPDF) with a maximum $Q$ of
35\,\AA\, was processed using the PDFGETN program.\cite{peterson_2000} Least squares
refinement of the NPDFs was performed using PDFGUI.\cite{farrow_2007} 

\section{Results and Discussion}
\subsection{Zn$_{1-x}$Co$_x$Cr$_2$O$_4$}

\zco\, and \cco\, are normal cubic spinels in the space group $Fd\overline{3}m$ at
room temperature. These systems vary significantly in their magnetic properties,
primarily due to the differences in properties of the $A$ site cations. Zn$^{2+}$
has the closed electron configuration [Ar]3$d^{10}$ which renders it magnetically
inert. Direct antiferromagnetic Cr$^{3+}$-Cr$^{3+}$ interactions in the pyrochlore
Cr sublattice of \zco\, give rise to geometric frustration. A spin-Jahn-Teller
distortion partially lifts spin frustration in \zco\, enabling the onset of
antiferromagnetic order at the N\'eel
temperature.\cite{kagomiya_2002,lee_2002,klemme_2004,lee_2007} The nuclear structure
of the spin-Jahn-Teller phase of \zco\, has been extensively
studied.\cite{kagomiya_2002,lee_2002,lee_2007,kemei_2013} Recently, our group has
proposed a structural model of coexisting tetragonal $I4_1/amd$ and orthorhombic
$Fddd$ phases for \zco.\cite{kemei_2013} On the other hand, Co$^{2+}$ has the
electronic configuration [Ar]3$d^7$ with three unpaired spins that interact
ferrimagnetically with Cr$^{3+}$ through Co$^{2+}$-O-Cr$^{3+}$ superexchange
interactions. \cco\, shows complex magnetic behavior; it undergoes a magnetic phase
transition from a paramagnetic state to a ferrimagnetic long-range ordered state
near 94\,K.\cite{Menyuk_1964,Tomiyasu_2004,lawes_2006}  A recent report by Chang
\textit{et al.} shows long range spiral order in \cco\, below 25\,K with an
incommensurate propagation vector and a transition at 14\,K to commensurate spiral
order.\cite{chang_2009} While no studies report a lowering of structural symmetry in
\cco, ultrasound propagation measurements performed on single crystals of \cco\, by
Tsurkan \textit{et al.} show a field-induced structural distortion to cubic symmetry
at high magnetic fields.\cite{tsurkan_2013} \cco\, exhibits spin charge coupling;
the onset of incommensurate spiral order in \cco\, is accompanied by a change in the
dielectric constant.\cite{lawes_2006,mufti_2010} In addition, the dielectric
constant shows magnetic field dependence below 95\,K.  \cite{mufti_2010} We examine
the changes in structural ground states of \zco\, when 10$\%$ and 20$\%$ Co$^{2+}$
cations are substituted on the non-magnetic Zn$^{2+}$ site.

\begin{figure}[h!]
\centering\includegraphics[width=8cm]{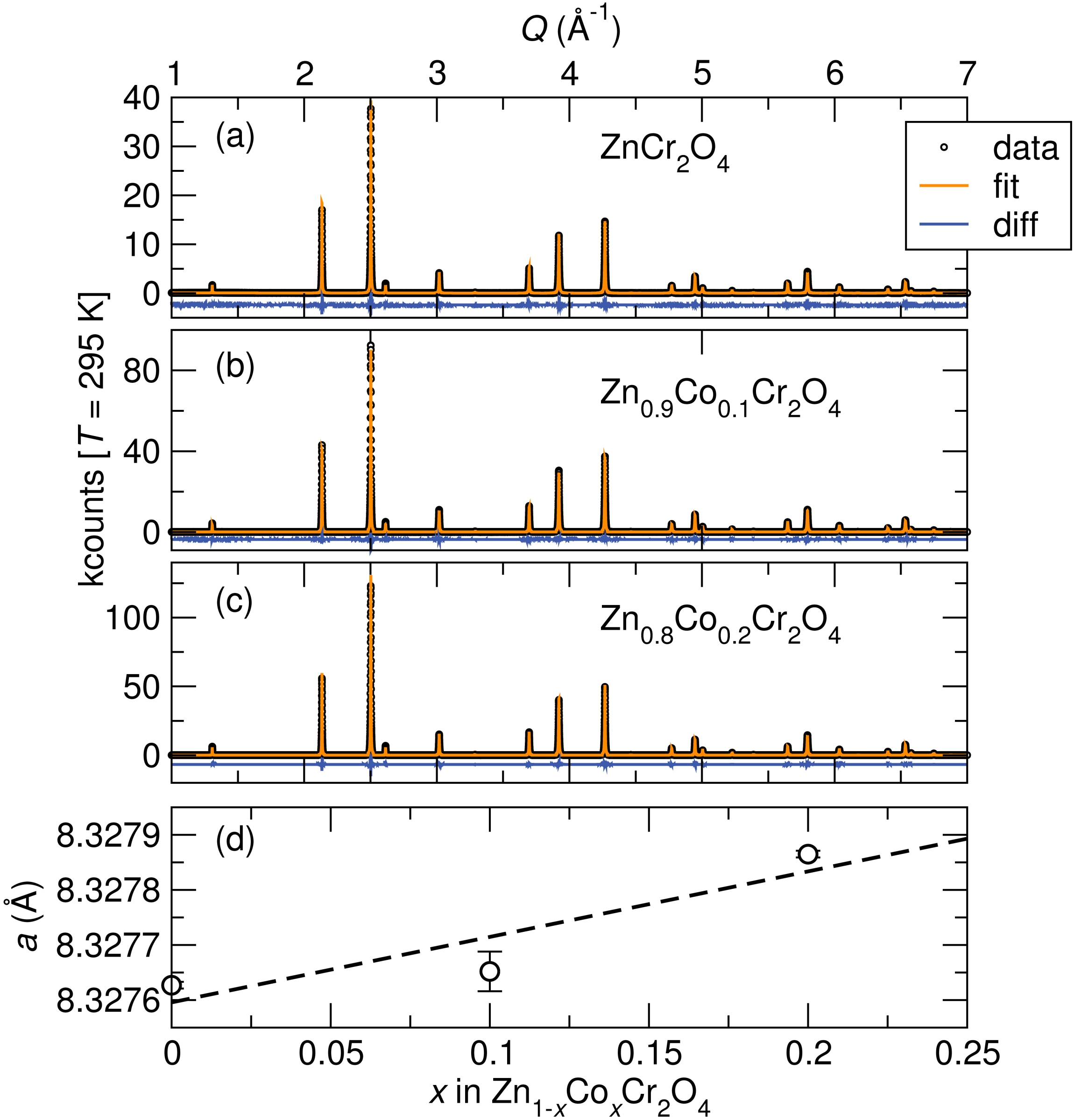}\\
\caption{(Color online) Room temperature synchrotron X-ray diffraction of the compounds (a) \zco, (b) Zn$_{0.9}$Co$_{0.1}$Cr$_2$O$_4$, and (c) Zn$_{0.8}$Co$_{0.2}$Cr$_2$O$_4$. All samples are well described in the cubic space group $Fd\overline{3}m$ and no impurity peaks are observed. (d) A unit cell expansion occurs with substitution of Co$^{2+}$ for Zn$^{2+}$ in Zn$_{1-x}$Co$_{x}$Cr$_2$O$_4$. The dashed line is a linear fit to the lattice parameters of Zn$_{1-x}$Co$_{x}$Cr$_2$O$_4$. } 
\label{fig:zccortstruct}
\end{figure}

\begin{table}[h]
\caption{\label{tab:magzcoco} 
Magnetic parameters of Zn$_{1-x}$Co$_x$Cr$_2$O$_4$.}
\centering
\begin{tabular}{lllllllllll}
\hline
&& $T_N$ (K)  &$\mu_{eff}$($\mu_B$) & $\mu_{calc}$($\mu_B$) & $\Theta_{CW}$(K)&  \ \\
\hline
\zco\, && 12.3 & 5.2 & 5.48 & -288 \\
Zn$_{0.9}$Co$_{0.1}$Cr$_2$O$_4$   && 9 & 5.4 & 5.61 & -350 \\
Zn$_{0.8}$Co$_{0.2}$Cr$_2$O$_4$   && 11 &5.6 & 5.74 & -369\\
\hline
\end{tabular}
\end{table}

At room temperature, the prepared compounds Zn$_{1-x}$Co$_x$Cr$_2$O$_4$ where
$x$\,$\leq$\,0.2 are cubic spinels in the space group $Fd\overline{3}m$ (Fig.
\ref{fig:zccortstruct} and Table \ref{tab:rietveldZnCoCr2O4} of the appendix). The
similarity in the tetrahedral ionic radii of Co$^{2+}$ and Zn$^{2+}$ allows the
entire solid solution Zn$_{1-x}$Co$_{x}$Cr$_2$O$_4$ to retain cubic
$Fd\overline{3}m$ symmetry at room temperature.\cite{melot_2009} Despite the smaller
ionic radius of Co$^{2+}$, a unit cell expansion occurs with substitution of
Co$^{2+}$ for Zn$^{2+}$ in Zn$_{1-x}$Co$_x$Cr$_2$O$_4$. The observed unit cell
expansion has been previously reported and is attributed to higher cation-cation
repulsion with increasing substitution of the more ionic Co$^{2+}$ for Zn$^{2+}$ in
Zn$_{1-x}$Co$_{x}$Cr$_2$O$_4$.\cite{melot_2009}  

\begin{figure*}
\centering\includegraphics[width=12cm]{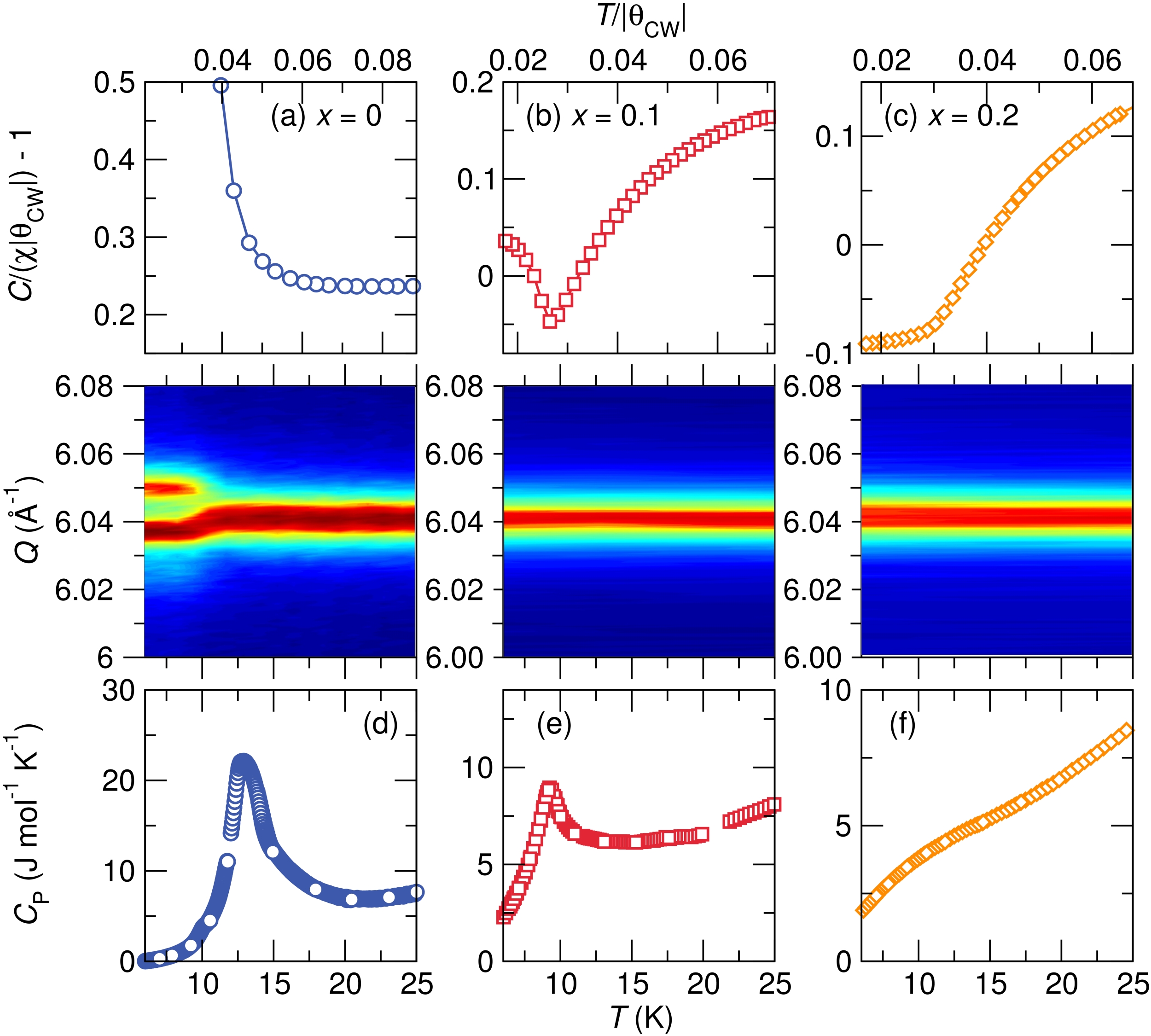}\\
\caption{(Color online) Suppression of the spin-Jahn-Teller distortion in Zn$_{1-x}$Co$_{x}$Cr$_2$O$_4$ with increase in $x$. (a) Inverse scaled magnetic susceptibility as a function of temperature for (a) \zco, (b) \zcoco, and (c) \zcococo\, measured in a 1000\,Oe field. Compensated antiferromagnetic interactions of \zco\, shown by the upward turn of the inverse susceptibility below the magnetic ordering temperature become uncompensated with the introduction of Co$^{2+}$ in place of Zn$^{2+}$ as illustrated by downward turn of the susceptibility of \zcococo\, below the N\'eel temperature.  The middle panel shows variable-temperature high-resolution X-ray powder diffraction of the cubic (800) reflection. Geometric frustration in \zco\, drives the lattice distortion shown by the splitting of the (800) reflection at the N\'eel temperature (12.3\,K) of \zco. The spin-Jahn-Teller distortion of \zco\, is suppressed even when only 10$\%$ Co$^{2+}$ cations are substituted for Zn$^{2+}$. (d) The sharp heat capacity anomaly observed at the spin-Jahn-Teller distortion temperature of \zco\, is suppressed in \zcoco\,(e) and strongly suppressed in \zcococo\,(f).} 
\label{fig:zccovt}
\end{figure*}

Antiferromagnetic interactions of the geometrically frustrated spinel \zco\,[Fig.
\ref{fig:zccovt} (a)] evolve to uncompensated antiferromagnetic interactions in
\zcococo\,[Fig. \ref{fig:zccovt} (c)]. Compensated interactions of \zco\, are
illustrated by the upward turn of the inverse susceptibility below the magnetic
ordering temperature and this contrasts with uncompensated antiferromagnetic
interactions in \zcococo\, where magnetic ordering yields a downward turn of the
inverse susceptibility. An increase in the number of magnetic interactions is
expected in Zn$_{1-x}$Co$_{x}$Cr$_2$O$_4$ with increase in $x$. Accordingly,
Curie-Weiss fitting in the paramagnetic regime (300\,$\leq\,T\,\leq$ 390\,K) of
Zn$_{1-x}$Co$_{x}$Cr$_2$O$_4$ yields an increasing $\Theta_{CW}$ with $x$ (Table
\ref{tab:magzcoco}). Similarly, the expected increase in effective moment with
Co$^{2+}$ substitution is also observed (Table \ref{tab:magzcoco}). When spins are
substituted on the non-magnetic $A$ sites of $A$Cr$_2$O$_4$ spinels at
concentrations greater than 20$\%$, spin frustration is
lifted.\cite{melot_2009,yan_2008,kemei_2012} However, at dilute $A$ site spin
concentrations, disorder in the spin interactions has been shown to further suppress
magnetic ordering. \cite{melot_2009,yan_2008,kemei_2012} Due to the disorder in spin
interactions, magnetic ordering in \zcoco\, and \zcococo\, occurs at lower
temperatures compared to \zco\,(Table \ref{tab:magzcoco}). 

\begin{figure}
\centering\includegraphics[width=8cm]{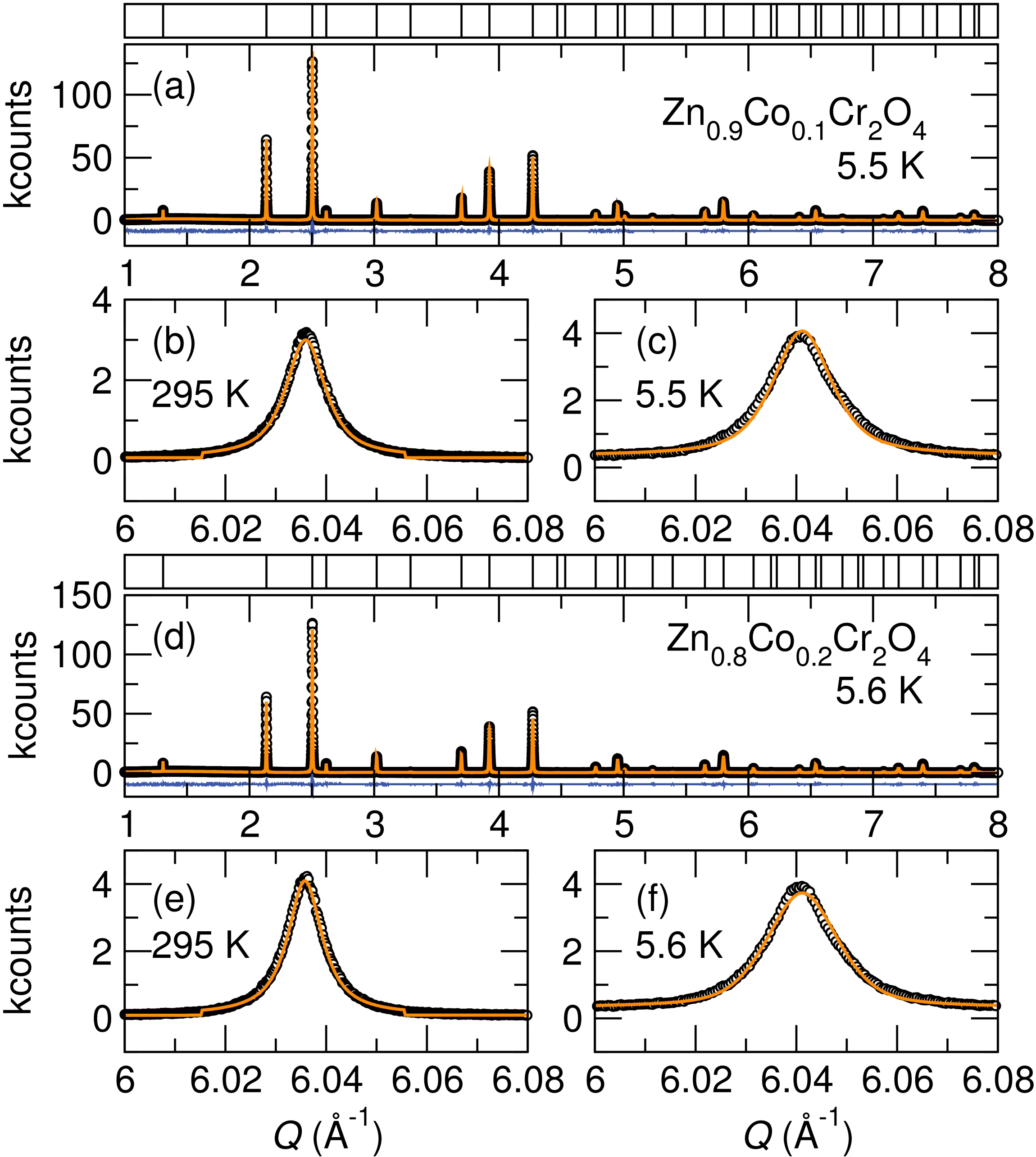}\\
\caption{(Color online) Low temperature synchrotron X-ray diffraction of (a) Zn$_{0.9}$Co$_{0.1}$Cr$_2$O$_4$, and (d) Zn$_{0.8}$Co$_{0.2}$Cr$_2$O$_4$ modeled to the cubic space group $Fd\overline{3}m$. The (800) reflections of  \zcoco\, and \zcococo\, at room temperature are shown here in (b) and (e) respectively and near 5\,K in (c) and (f) respectively. A broadening of the (800) reflection is observed at low temperature in \zcoco\, and \zcococo.} 
\label{fig:zccolt}
\end{figure}

Antiferromagnetic order in \zco\, occurs concurrently with a structural distortion
(middle panel of Fig. \ref{fig:zccovt}).  The structural ground state of \zco\, has
been extensively investigated and a recent report from our group shows that the
spin-Jahn-Teller phase of \zco\, is best described by the combination of tetragonal
$I4_1/amd$ and orthorhombic $Fddd$ space
groups.\cite{kemei_2013,lee_2007,kagomiya_2002} While \zco\, exhibits a clear
lattice distortion at the N\'eel temperature, the cubic $Fd\overline{3}m$ (800)
reflection for samples $x$\,=\,0.1 and $x$\,=\,0.2 shows no divergence illustrating
the complete suppression of long-range structural distortion in these materials. As
a result, the average nuclear structures of \zcoco\, and \zcococo\, near 5\,K are
well modeled by the cubic space group $Fd\overline{3}m$ as illustrated in Fig.
\ref{fig:zccolt}. However, a clear peak broadening of the (800) reflection occurs
near 5\,K in \zcoco\, and \zcococo\, as shown in Fig. \ref{fig:zccolt} (c) and (f)
respectively. This broadening is indicative of higher strain at low temperatures
that can result from local distortions in these materials.  The structural
parameters of \zcoco\, and \zcococo\, are tabulated in Table
\ref{tab:rietveldZnCoCr2O4} of the appendix.

The spin-Jahn-Teller distortion of \zco\, yields a sharp heat capacity anomaly at
$T_N$ [Fig. \ref{fig:zccovt} (d)]. This heat capacity anomaly is slightly suppressed
in \zcoco\, and it becomes very broad in \zcococo\,[Fig. \ref{fig:zccovt} (e) and
(f)]. The suppression of the heat capacity anomalies in \zcoco\, and \zcococo\,
shows that these systems host residual spin and structural disorder. 

We have shown the differences in structural ground state when $\geq$\,10$\%$ of
Co$^{2+}$ are substituted on the non-magnetic $A$ site of \zco. The structural
distortion that accompanies magnetic ordering in \zco\, is completely suppressed
even for only 10$\%$ Co$^{2+}$ substitution in \zcoco. This suggests that random
Co-O-Cr superexchange interactions in Zn$_{1-x}$Co$_{x}$Cr$_2$O$_4$, partially break
the spin ground state degeneracy of these systems, allowing the onset of a magnetic
ground state without the need for a long range structural distortion. It is also
plausible that random Co-O-Cr superexchange interactions could be disrupting the
coherency of Cr-Cr exchange coupling paths thus inhibiting spin-Jahn-Teller
distortions in \zcoco\, and \zcococo. Small substitutions of Co$^{2+}$ for Mg$^{2+}$
will likely suppress the Spin-Jahn-Teller distortion of \mco; this is supported by
the similar structural effects of Cu$^{2+}$ substitutions for Mg$^{2+}$ and
Zn$^{2+}$ in \mco\, and \zco\, as discussed in the following sections. 

\subsection{Mg$_{1-x}$Cu$_x$Cr$_2$O$_4$}
\begin{figure}[!h]
\centering\includegraphics[width=8cm]{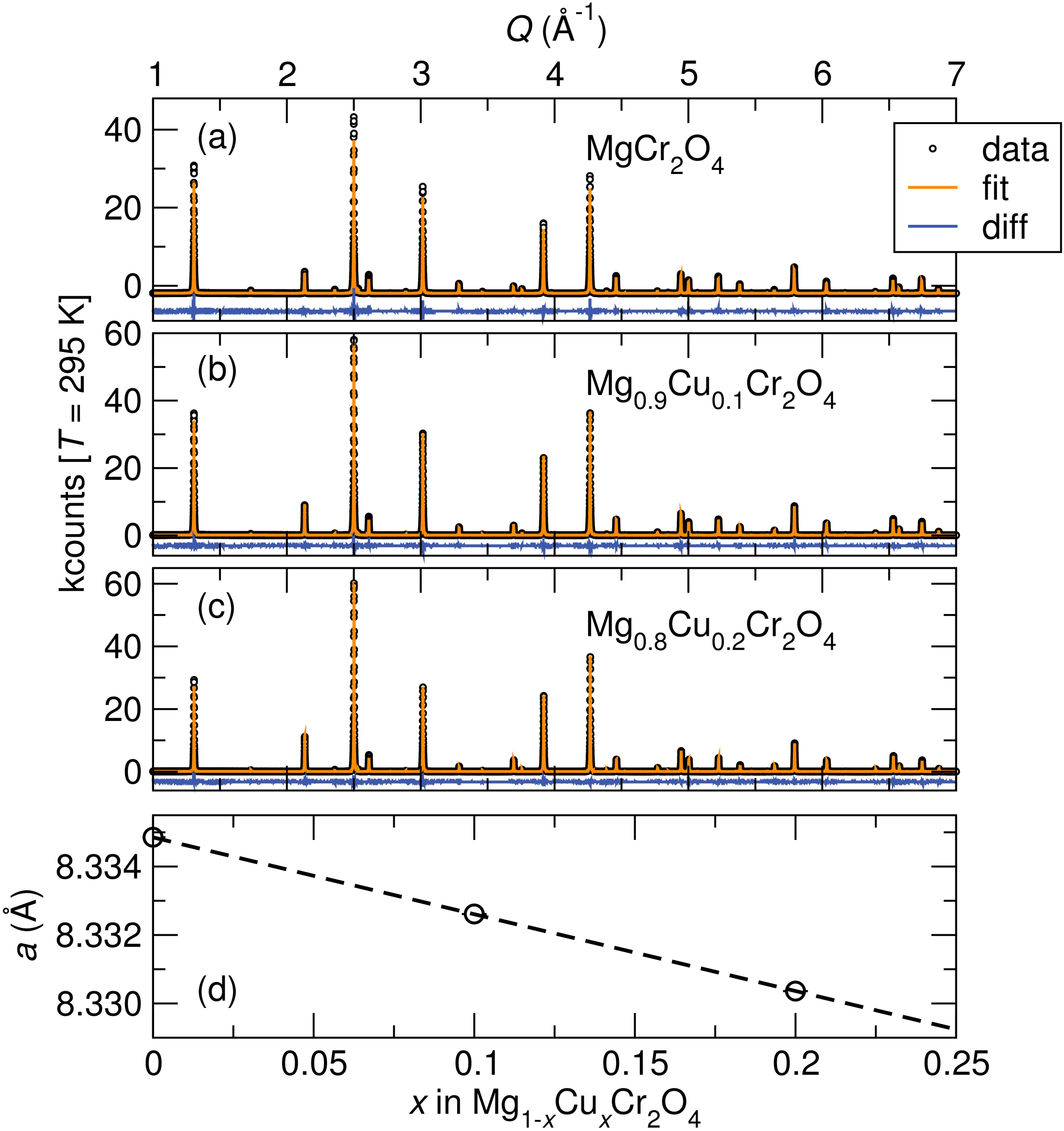}\\
\caption{(Color online) Rietveld refinement of room temperature high resolution synchrotron X-ray powder diffraction of (a) \mco\, (b) \mcuco\, and (c) \mcucuco\, to the cubic space group $Fd\overline{3}m$. All samples show a very small Cr$_2$O$_3$ impurity with concentrations $<$\,1$\%$ in all samples. (d) The cubic lattice constant in Mg$_{1-x}$Cu$_x$Cr$_2$O$_4$ decreases linearly with increase in Cu$^{2+}$ concentration. Error bars are smaller than data symbols.} 
\label{fig:mccort}
\end{figure}

We examine the effect of both spin and lattice disorder on the spin-Jahn-Teller
ground state of \mco\, by substituting $\geq$\,10$\%$ of Jahn-Teller active
Cu$^{2+}$ for Mg$^{2+}$. At room temperature the normal spinels \mco\, and \cuco\,
have different structural ground states; \mco\, is cubic while \cuco\, is
tetragonal. The tetragonal structure of \cuco\, results from cooperative Jahn-Teller
ordering of CuO$_4$ tetrahedra at $T$\,=\,853\,K.\cite{ye_1994} \mco\, is a
frustrated antiferromagnet, and its transition to an ordered magnetic state at
$T_N$\,=\,12.9\,K is accompanied by a structural
distortion.\cite{martin_2008,ehrenberg_2002} The spin-Jahn-Teller distorted phase of
\mco\, had been previously described by the tetragonal $I4_1/amd$
structure,\cite{martin_2008,ehrenberg_2002}  but we have recently shown that this
system consists of coexisting tetragonal $I4_1/amd$ and orthorhombic $Fddd$
phases.\cite{kemei_2013} \cuco\, is ferrimagnetic, with magnetic Cu$^{2+}$ and
Cr$^{3+}$ sublattices contributing to a non-collinear magnetic structure where two
canted Cr$^{3+}$ sublattices yield a magnetic moment that is partially compensated
by the Cu$^{2+}$ sublattice at $T_N$\,=\,135\,K.\cite{prince_1957} In addition to
the high temperature Jahn-Teller driven cubic-tetragonal phase transition, \cuco\,
undergoes yet another structural distortion from tetragonal $I4_1/amd$ to
orthorhombic $Fddd$ symmetry near 130$\,$K due to magnetostructural
coupling.\cite{suchomel_2012} 

The prepared spinel solid solutions Mg$_{1-x}$Cu$_x$Cr$_2$O$_4$ where
$x$\,$\leq$\,0.2 are cubic with the space group $Fd\overline{3}m$ at room
temperature [Fig. \ref{fig:mccort} (a),   (b), and (c)]. While tetrahedral Mg$^{2+}$
and Cu$^{2+}$ have identical Shannon-Prewitt ionic radii, we observe a lattice
contraction with increase in Cu$^{2+}$, following Vegard's law [Figure
\ref{fig:mccort} (d)]. This decrease in lattice constant with Cu$^{2+}$ substitution
is consistent with the earlier work by Shoemaker and Seshadri\cite{shoemaker_2010}
and with the smaller pseudocubic cell volume of
\cuco\,(566.38\,\AA$^3$)\cite{suchomel_2012} compared with that of
\mco(579.017\,\AA) at room temperature.\cite{kemei_2013} The structural parameters
of Mg$_{1-x}$Cu$_x$Cr$_2$O$_4$ for $x$ = 0.1 and 0.2 at 300\,K are tabulated in
Table. \ref{tab:rietveldmcco} of the appendix.

\begin{table}
\caption{\label{tab:magmcuco} 
Magnetic parameters of Mg$_{1-x}$Cu$_x$Cr$_2$O$_4$.}
\centering
\begin{tabular}{lllllllll}
\hline
&& $T_N$ (K) & $\mu_{exp}$($\mu_B$) & $\mu_{calc}$($\mu_B$) & $\Theta_{CW}$(K)&  \ \\
\hline
\mco\, &&  12.9 & 5.2 & 5.47 & -368 &&\\
\mcuco\, &&  11 & 5.3 & 5.50 & -361 &\\
\mcucuco\, && 15 &5.2  & 5.53 & -329 &\\
\hline
\end{tabular}
\end{table}

\begin{figure*}
\centering\includegraphics[width=12cm]{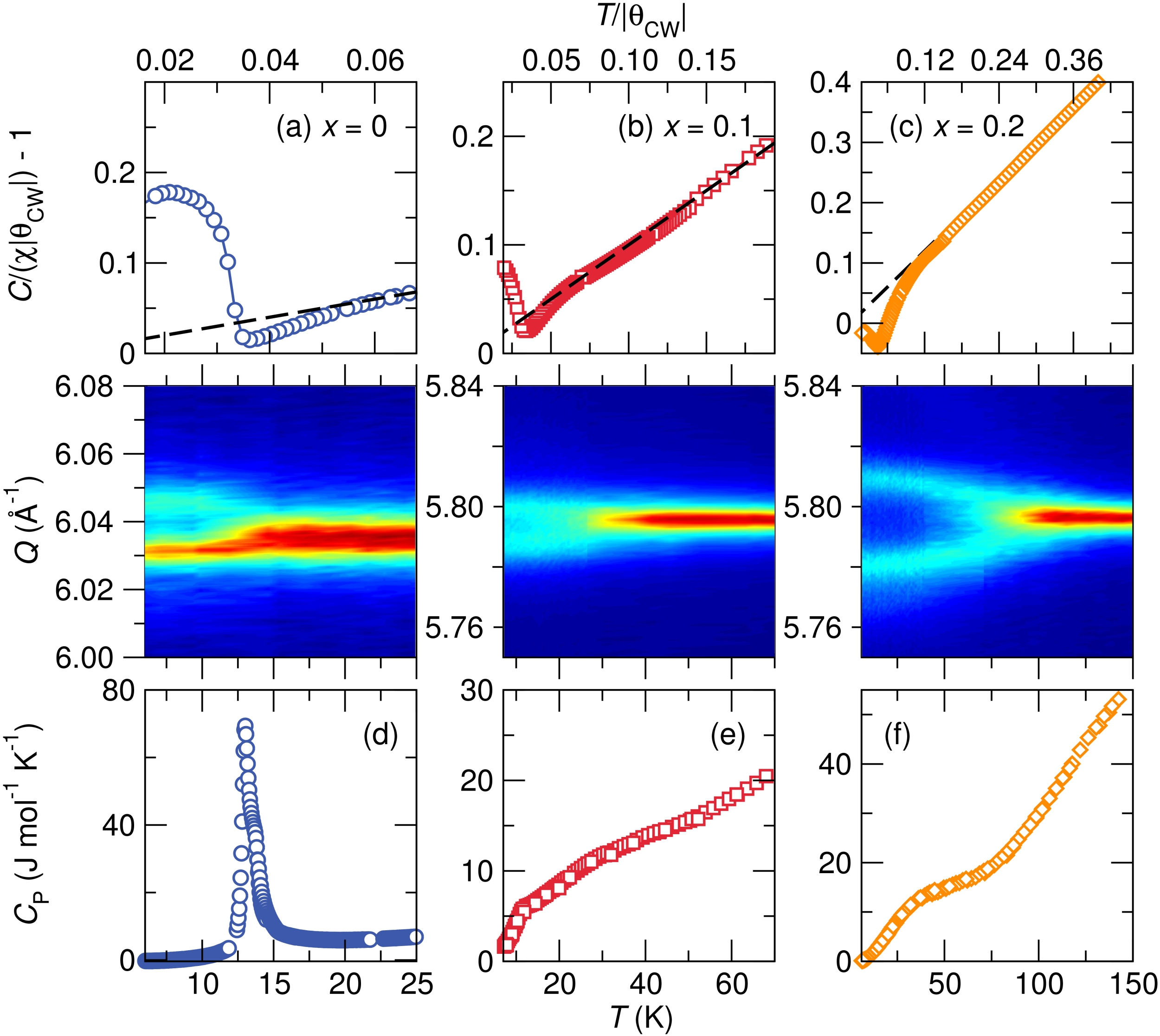}\\
\caption{(Color online) Spin-Jahn-Teller and Jahn-Teller ordering in Mg$_{1-x}$Cu$_x$Cr$_2$O$_4$. The top panel shows the scaled inverse susceptibility of \mco\,(a), \mcuco\,(b) and \mcucuco\,(c) measured in a 1000\,Oe field. Compensated antiferromagnetic interactions in \mco\,(a) and \mcuco\,(b) evolve to uncompensated antiferromagnetic interactions in \mcucuco\,(c). Geometric frustration of spins in \mco\, drives a structural distortion at $T_N$\,=\,12.9\,K that is indicated by the splitting of the high symmetry (800) reflection. Cooperative Jahn-Teller ordering spurs average structure distortions in \mcuco\, at $T$\,$\sim$\,35\,K and \mcucuco\, at $T$\,$\sim$\,110\,K. The structural distortions in \mcuco\, and \mcucuco\, are decoupled from the magnetism and no further structural distortions occur near the N\'eel temperature of these compounds although they exhibit spin frustration. (d) There is a sharp heat capacity anomaly at the N\'eel temperature of \mco\, with a shoulder feature plausibly indicating a slight separation in temperature of the magnetic and structural transitions. (e) \mcuco\, shows a broad heat capacity anomaly with a kink at $T_N$. (f) Similarly, \mcucuco\, shows a broad heat capacity peak in the temperature range 6\,K$\leq$\,$T$\,$\leq$80\,K. } 
\label{fig:mccovt}
\end{figure*}

The evolution of magnetism in Mg$_{1-x}$Cu$_x$Cr$_2$O$_4$ where $x$\,$\leq$\,0.2, is
similar to that observed in Zn$_{1-x}$Co$_x$Cr$_2$O$_4$; frustrated
antiferromagnetism in \mco\, evolves to uncompensated antiferromagnetism in \mcucuco
[Fig. \ref{fig:mccovt} (a), (b), and (c)]. Antiferromagnetic ordering in \mcuco\,
occurs at a lower temperature than in \mco\, due to dilute $J_{Cu-O-Cr}$ couplings
interfering with $J_{Cr-Cr}$ couplings (Table \ref{tab:magmcuco}). However, the
increase in Cu$^{2+}$ concentration in \mcucuco\, yields a higher magnetic ordering
temperature and this is consistent with the findings that sufficient magnetic $A$
site spins lift frustration in geometrically frustrated $A$Cr$_2$O$_4$
spinels.\cite{melot_2009, kemei_2012, yan_2008} Curie-Weiss fitting in the
paramagnetic regime $300\,K\,\leq\,T\,\leq\,390\,K$, of Mg$_{1-x}$Cu$_x$Cr$_2$O$_4$
yields a slight increase in the effective moment of \mcuco\, and a weakening of the
overall strength of magnetic interactions shown by the decrease in the magnitude of
$\Theta_{CW}$(Table \ref{tab:magmcuco}). Weaker antiferromagnetic interactions with
Cu$^{2+}$ substitution are attributed to the effects of spin disorder due to dilute
$A$ site spins. The decrease in $\Theta_{CW}$ in Mg$_{1-x}$Cu$_x$Cr$_2$O$_4$ with
increase in $x$ contrasts with the increase in $\Theta_{CW}$ in
Zn$_{1-x}$Co$_x$Cr$_2$O$_4$ with increase in $x$; this difference is attributed to
the higher spin of Co$^{2+}$ 3$d^7$ $s$\,=\,$\frac{3}{2}$ compared to Cu$^{2+}$
3$d^9$ $s$\,=\,$\frac{1}{2}$. The higher spin of Co$^{2+}$ contributes to stronger
magnetic interactions in Zn$_{1-x}$Co$_x$Cr$_2$O$_4$.
 
\begin{figure}
\centering\includegraphics[width=8cm]{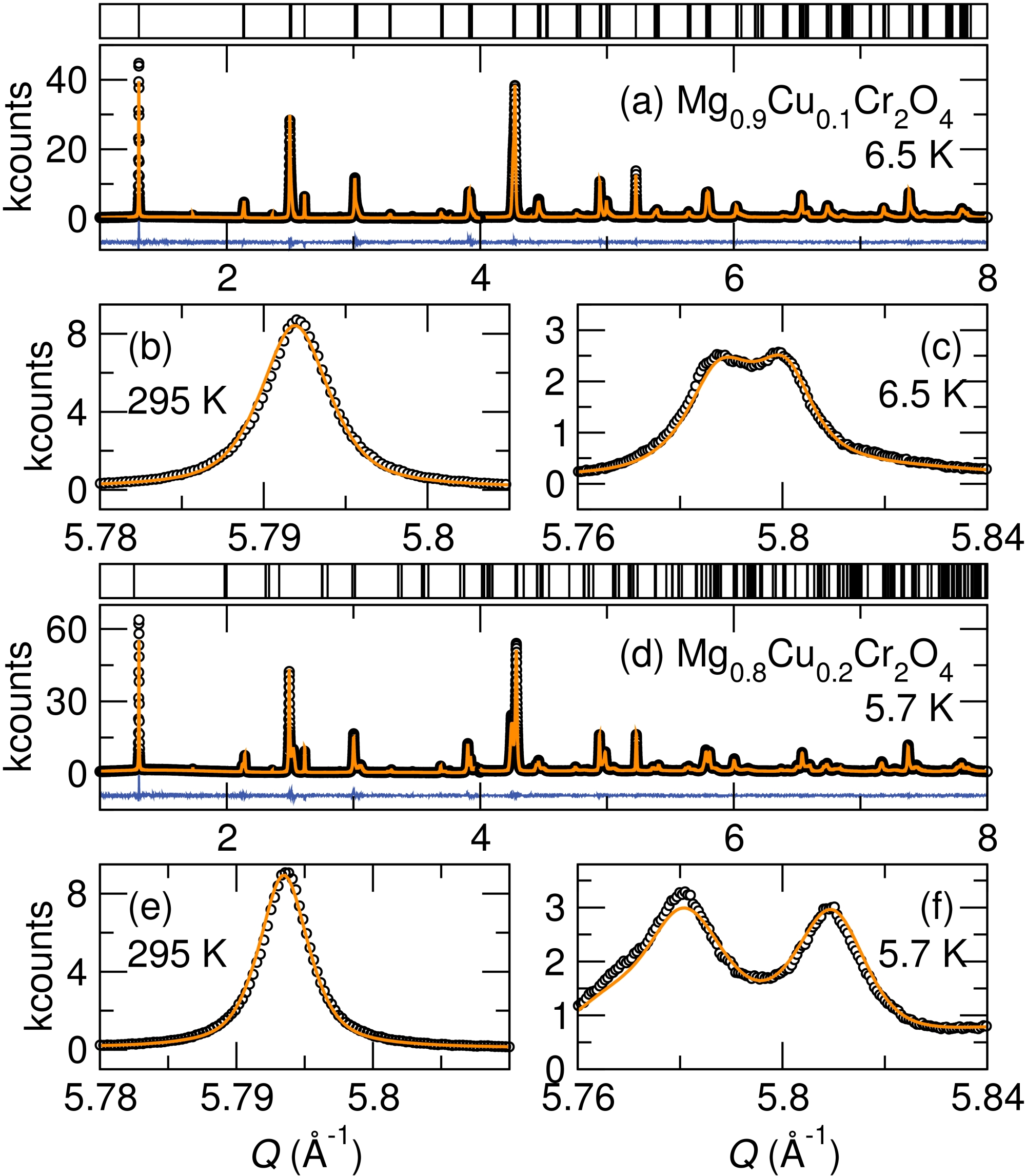}\\
\caption{(Color online) Low temperature structures of  (a) \mcuco\, and (d) \mcucuco\, indexed to the orthorhombic $Fddd$ structure. Data is shown in black, the structural model in orange and the difference between the structural model and the data is in blue. Below the respective Jahn-Teller ordering temperatures of \mcuco\, and \mcucuco, the coincident (731) and (553) reflections shown in (b) and (c) split into several reflections as shown in (c) and (f).} 
\label{fig:mccolt}
\end{figure}

Structural distortions are observed in all compounds Mg$_{1-x}$Cu$_x$Cr$_2$O$_4$
when $x$\,$\leq$\,0.2. The spin-Jahn-Teller distortion of \mco\, is illustrated by
the splitting of the (800) reflection to several low temperature peaks (Leftmost
middle panel of Fig. \ref{fig:mccovt}). While degeneracy in spin ground states
drives the structural distortion in \mco, degeneracy in the orbital configurations
of tetrahedral Cu$^{2+}$ drive Jahn-Teller distortions in \mcuco\, and \mcucuco\, 
at 35\,K and 110\,K respectively. The middle panel of figure \ref{fig:mccovt} shows
the splitting of the coincident (731) and (553) reflections in \mcuco\, and
\mcucuco. The Jahn-Teller distortion increases with Cu$^{2+}$ concentration as shown
by the larger separation between the low-temperature reflections of \mcucuco\, and
the onset of this distortion at higher temperature in this compound. We note that
magnetic transitions do not accompany the structural distortions of \mcuco\, and
\mcucuco. The Jahn-Teller phases of \mcuco\, and \mcucuco\ are well modeled by the
orthorhombic $Fddd$ structure (Fig. \ref{fig:mccolt}) that is ascribed to \cuco\,
following its magnetostructural distortion.\cite{suchomel_2012} An important
difference between \cuco\, and these systems studied here is that the orthorhombic
structure of \cuco\, occurs due to magnetostructural coupling\cite{suchomel_2012}
while the orthorhombic structure of \mcuco\, and \mcucuco\, occurs in the
paramagnetic regime driven primarily by cooperative Jahn-Teller ordering. 

The spin-Jahn-Teller distortion of \mco\, results in a sharp heat capacity anomaly
with a slight shoulder feature [Fig. \ref{fig:mccovt} (d)]. The shoulder feature is
likely due to a slight separation in temperature of the magnetic and structural
changes. The onset of heat capacity changes in \mcuco\, and \mcucuco\, occur at high
temperatures where structural changes begin and they persist to low temperatures
where magnetic ordering occurs [Fig. \ref{fig:mccovt} (e) and (f)]. 

While the substitution of Co$^{2+}$ for Zn$^{2+}$ suppresses spin-Jahn-Teller
distortion in \zco, $\geq$\,10$\%$ substitution of Jahn-Teller active Cu$^{2+}$ for
Mg$^{2+}$ induces structural distortions in \mcuco\, and \mcucuco\, at temperatures
above the magnetic ordering temperatures of these compounds. The structural
distortions in \mcuco\, and \mcucuco\, affect the pyrochlore Cr sublattice; while
there is only one Cr-Cr bond length at room temperature in the cubic phases of these
systems, there are three Cr-Cr bond lengths in the orthorhombic phases of these
materials. Surprisingly, spin interactions remain frustrated in \mcuco\, and
\mcucuco\, with magnetic ordering occurring below 18\,K despite the presence of
distortions in the pyrochlore Cr sublattice of these materials at temperatures above
their N\'eel temperatures. No further structural distortions are observed in
\mcuco\, and \mcucuco\ near the N\'eel temperature. 

\subsection{Zn$_{1-x}$Cu$_x$Cr$_2$O$_4$}
\begin{figure}[!h]
\centering\includegraphics[width=8cm]{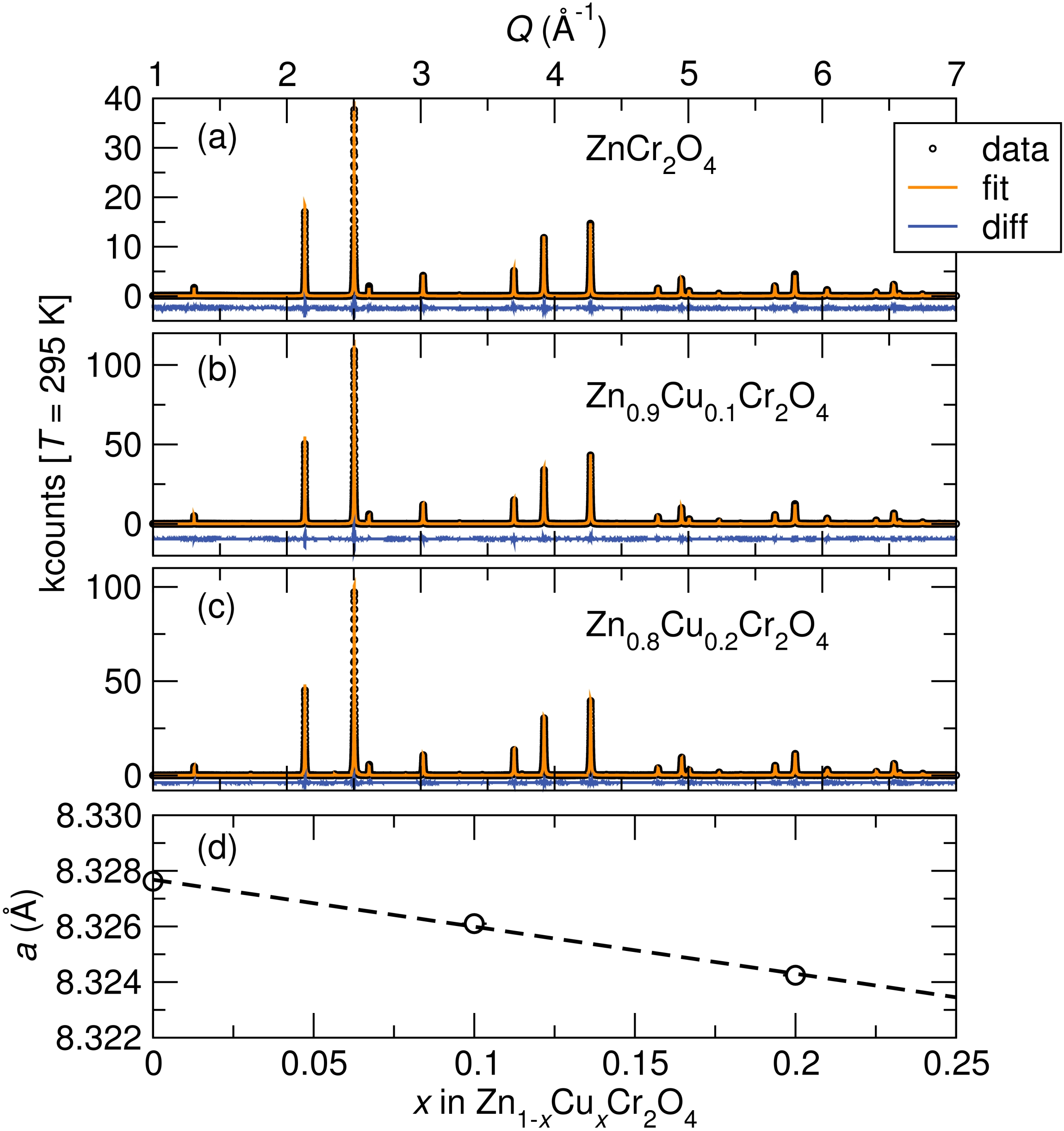}\\
\caption{(Color online) High resolution synchrotron X-ray powder diffraction of the systems (a) \zco\, (b) \zcuco\, and (c) \zcucuco\, measured at room temperature. All compounds are well indexed by the cubic space group $Fd\overline{3}m$. (d) A linear decrease of the cubic lattice constant occurs with Cu$^{2+}$ substitution for Zn$^{2+}$. Error bars are smaller than data symbols.} 
\label{fig:zcucort}
\end{figure}

We examine the effect of Cu$^{2+}$ substitutions for Zn$^{2+}$ on the
spin-Jahn-Teller distortion of \zco.  All prepared samples
Zn$_{1-x}$Cu$_x$Cr$_2$O$_4$ where $x$\,$\leq$\,0.2 are cubic spinels in the space
group $Fd\overline{3}m$ at room temperature as shown in Fig \ref{fig:zcucort}. Like
in the solid solutions Mg$_{1-x}$Cu$_x$Cr$_2$O$_4$, the substitution of Cu$^{2+}$
for Zn$^{2+}$ results in a steady decrease of the lattice constant [Fig
\ref{fig:zcucort} (d)]. This lattice decrease is in good agreement with the smaller
cell volume of \cuco(566.38\,\AA$^3$)\cite{suchomel_2012} at room temperature
compared to \zco(577.520\,\AA$^3$).\cite{kemei_2013} The structural parameters of
Zn$_{1-x}$Cu$_x$Cr$_2$O$_4$ are tabulated in Table \ref{tab:rietveldzcuco} of the
appendix.

\begin{figure*}
\centering\includegraphics[width=12cm]{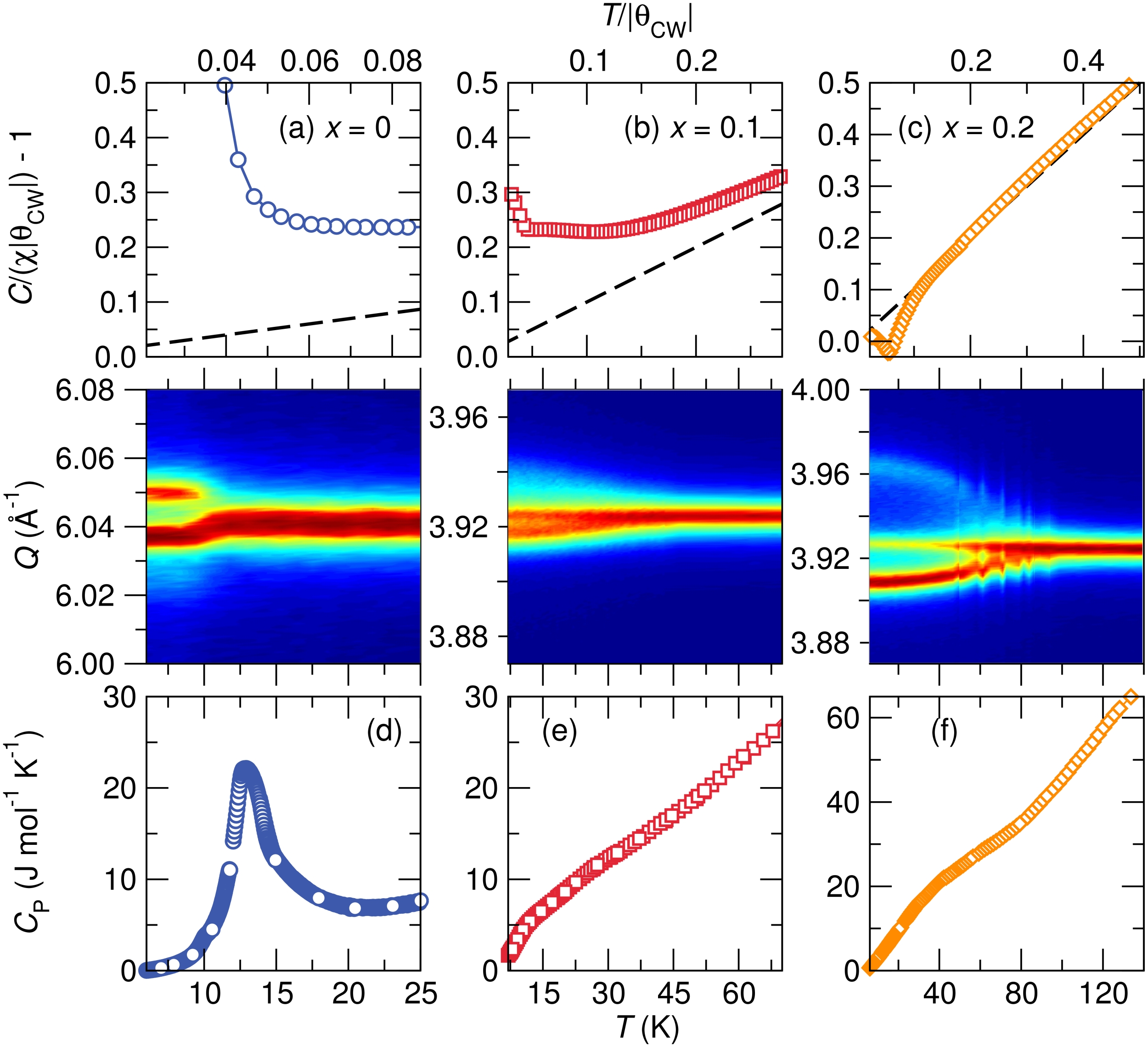}\\
\caption{(Color online) Spin-Jahn Teller and Jahn-Teller distortions in Zn$_{1-x}$Cu$_x$Cr$_2$O$_4$. The top panel shows inverse scaled susceptibility measurements of Zn$_{1-x}$Cu$_x$Cr$_2$O$_4$ measured under a 1000\,Oe field. Compensated antiferromagnetism is observed in \zco\,(a) and \zcuco\,(b) below the N\'eel temperature while \zcucuco\,(c) shows uncompensated antiferromagnetism. A lattice distortion accompanies magnetic ordering in \zco\, as shown by the splitting of the high symmetry (800) reflection at the N\'eel temperature. Jahn-Teller active Cu$^{2+}$ on the $A$ sites of \zcuco\, and \zcucuco\, drive lattice distortions at approximately 45\,K and 110\,K respectively, where the coincident (511) and (333) reflections split into several low temperature reflections. There is a large heat capacity anomaly at the spin-Jahn-Teller distortion temperature of \zco\, (d). Broad heat capacity anomalies are observed in \zcuco\,(e)  and \zcucuco\,(f) over the temperature range where structural and magnetic changes occur. The line features in the variable temperature data of sample \zcucuco\, are due to slight temperature fluctuations during the measurement.} 
\label{fig:zcucovt}
\end{figure*}

Compensated antiferromagnetic interactions in \zco\, and \zcuco\, evolve to
uncompensated antiferromagnetic interactions in \zcucuco\, 
[Figure \ref{fig:zcucovt}(a), (b), and (c)]. 
The onset of magnetic ordering in \zcuco\, occurs at lower
temperatures than in \zco\, due to disorder arising from dilute $A$ site spins while
\zcucuco\, shows the highest ordering temperature of the studied
Zn$_{1-x}$Cu$_x$Cr$_2$O$_4$ compounds (Table \ref{tab:magzcuco}). A slight decrease
in the effective moment is observed with Cu$^{2+}$ substitution in \zco\, and this
is attributed to the presence of short range spin interactions in the paramagnetic
regime contributing to the underestimation of the effective moment (Table
\ref{tab:magzcuco}). As observed in Mg$_{1-x}$Cu$_x$Cr$_2$O$_4$, a decrease in the
magnitude of $\Theta_{CW}$ occurs in Zn$_{1-x}$Cu$_x$Cr$_2$O$_4$ with increase in
$x$ suggesting that dilute Cu$^{2+}$ substitutions weaken the overall strength of
magnetic interactions in \mco\, and \zco. 

\begin{table}[h]
\caption{\label{tab:magzcuco} 
Magnetic parameters of Zn$_{1-x}$Cu$_x$Cr$_2$O$_4$.}
\centering
\begin{tabular}{lllllllll}
\hline
&& $T_N$ (K) & $\mu_{exp}$($\mu_B$) & $\mu_{calc}$($\mu_B$) & $\Theta_{CW}$(K)&  \ \\
\hline
\zco\, &&  12.3 & 5.2 & 5.47 & -288 &&\\
\zcuco\, &&  11 & 4.9 & 5.50 & -239 &\\
\zcucuco\, && 16 &5.0  & 5.53 & -270 &\\
\hline
\end{tabular}
\end{table}

\begin{figure}
\centering\includegraphics[width=9cm]{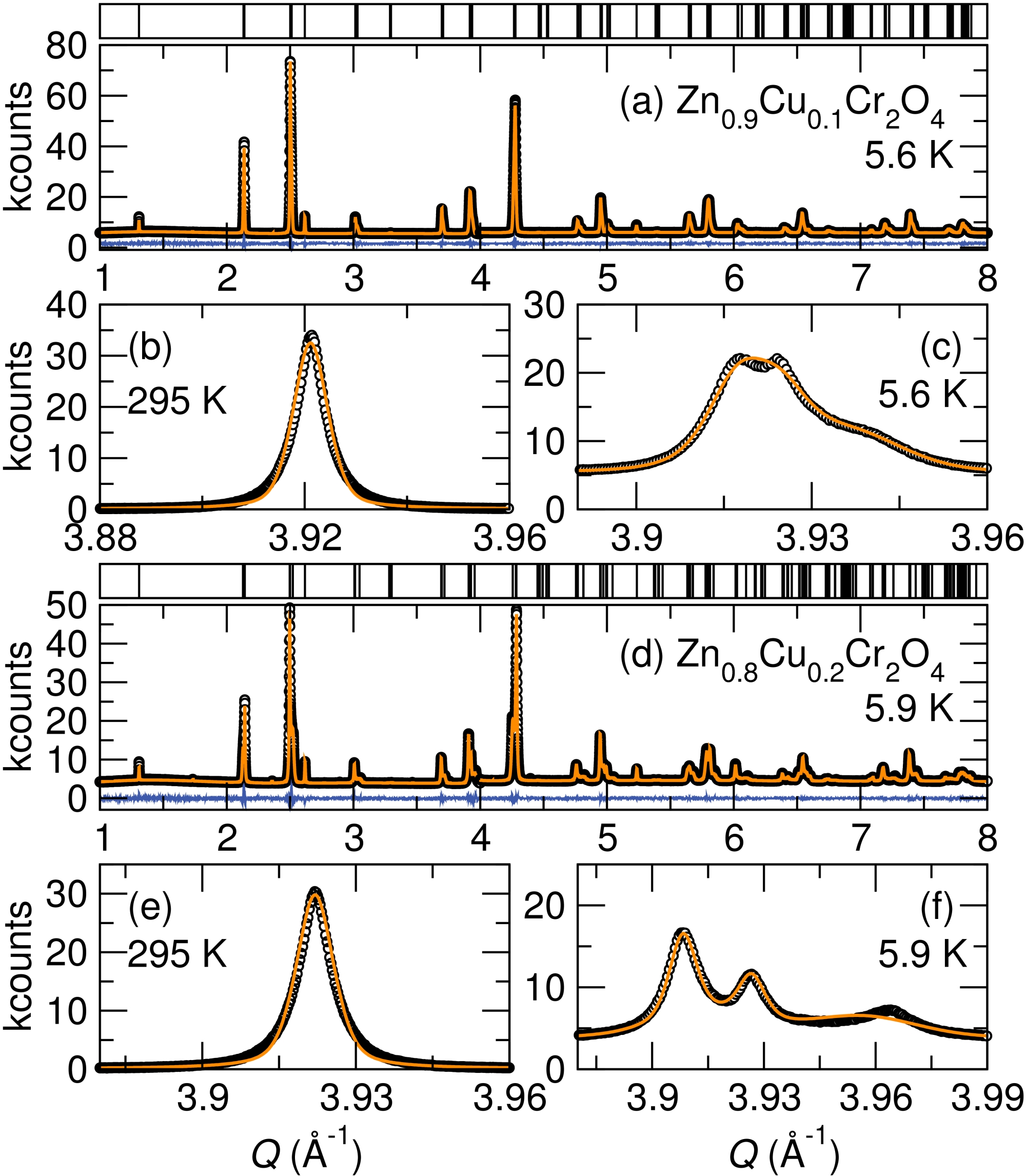}\\
\caption{(Color online) Synchrotron powder diffraction patterns of \zcuco\,(a) and \zcucuco\,(d) collected near 6\,K and indexed to the orthorhombic space group $Fddd$. Data is shown in black, the model is in orange while the difference is in blue. The high temperature coincident cubic $Fd\overline{3}m$ reflections (511) and (333) shown in (b) for \zcuco\, and in (e) for \zcucuco\, are split at lower temperatures following the cubic to orthorhombic lattice distortion. These low temperature reflections as indexed to the orthorhombic $Fddd$ structure are shown in (c) and (f) for \zcuco\, and \zcucuco\, respectively.} 
\label{fig:zcucolt}
\end{figure}

\begin{figure}
\centering\includegraphics[width=9cm]{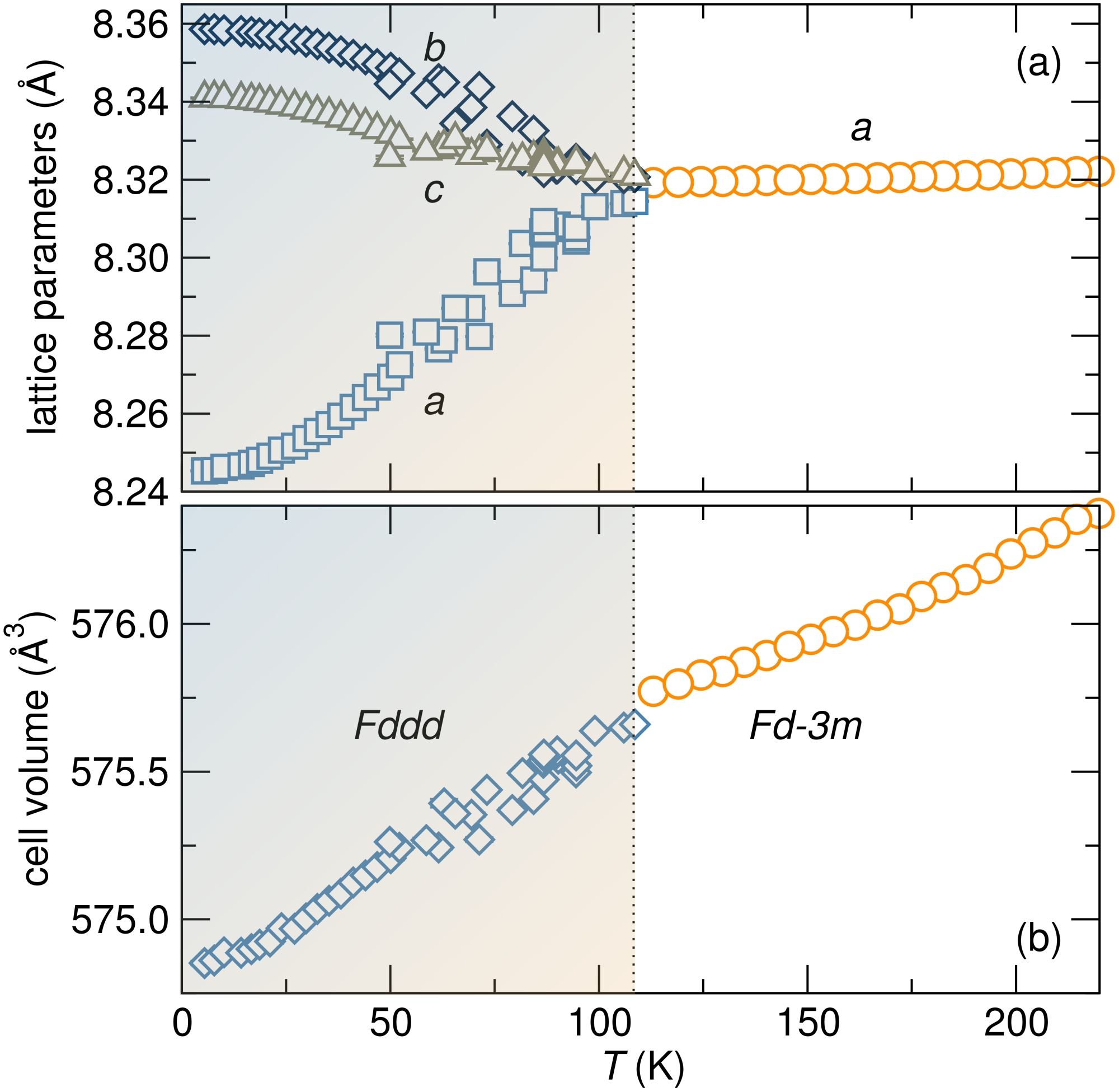}\\
\caption{(Color online) (a) The evolution of lattice parameters of \zcucuco\, as a function of temperature revealing a structural distortion at 110\,K where three orthorhombic $Fddd$ lattice constants emerge from the cubic $Fd\overline{3}m$ lattice constant. (b) There is a slight change in slope in the temperature dependent cell volume of \zcucuco\, at the structural distortion temperature. Error bars are smaller than data symbols.} 
\label{fig:lattice}
\end{figure}

\begin{figure} [h!]
\centering\includegraphics[width=9.5cm]{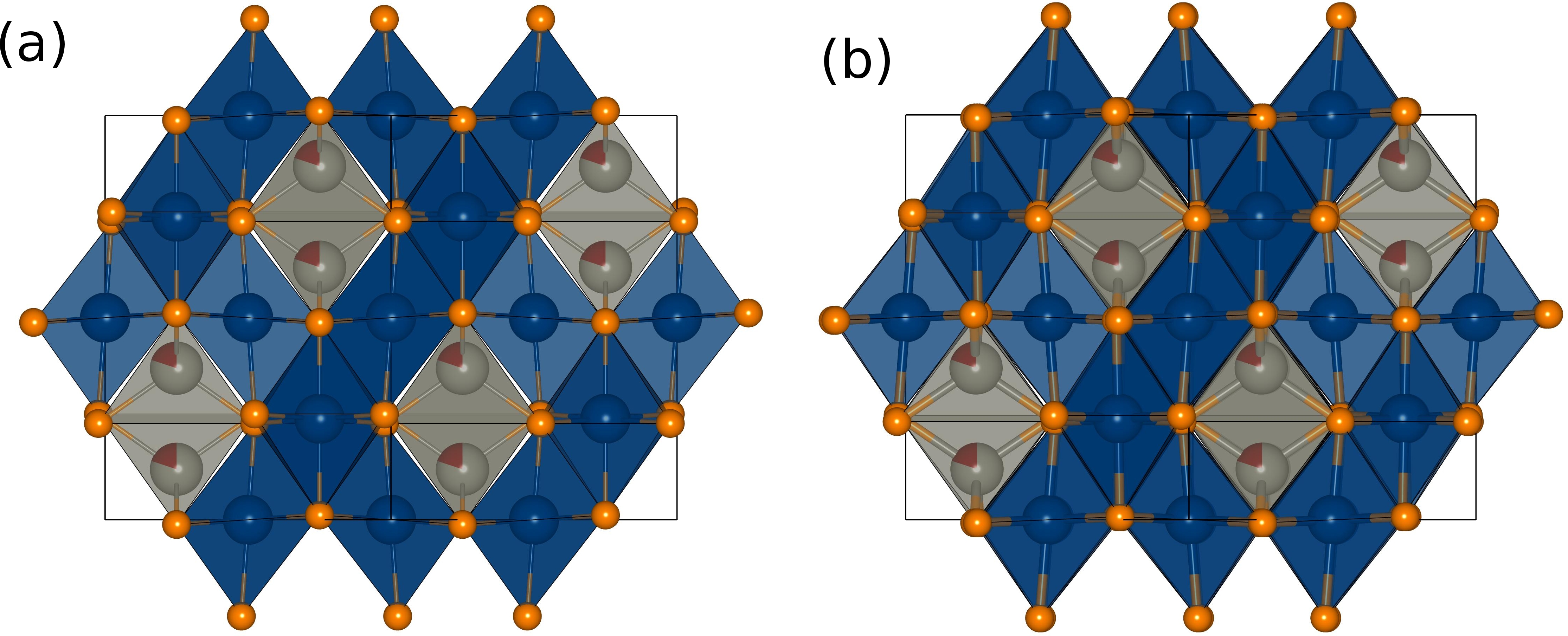}\\
\caption{(Color online) The spinel structure of \zcucuco\, in the cubic $Fd\overline{3}m$ phase near 300\,K and in the orthorhombic $Fddd$ phase near 6\,K are shown in (a) and (b) respectively. Edge sharing CrO$_6$(blue) octahedra are corner connected to (Zn/Cu)O$_4$(grey) tetrahedra. The shared (Zn/Cu) atomic site is shown in grey(Zn atomic fraction) and dark red(Cu atomic fraction). The ideal tetrahedral angle of 109.54$^{\circ}$ observed in the cubic phase(a) of \zcucuco\, is distorted to two angles of 111.898$^{\circ}$ and 105.99$^{\circ}$ in the orthorhombic phase(b); the (Zn/Cu)O$_4$(grey) tetrahedra appear more flattened in the orthorhombic phase(b) filling the tetrahedral voids between the CrO$_6$ octahedra while small gaps can be seen between the(Zn/Cu)O$_4$(grey) tetrahedra and the CrO$_6$(blue) octahedra in the cubic phase(a) of \zcucuco.}
\label{fig:vesta}
\end{figure} 

\begin{figure} [h!]
\centering\includegraphics[width=8.2cm]{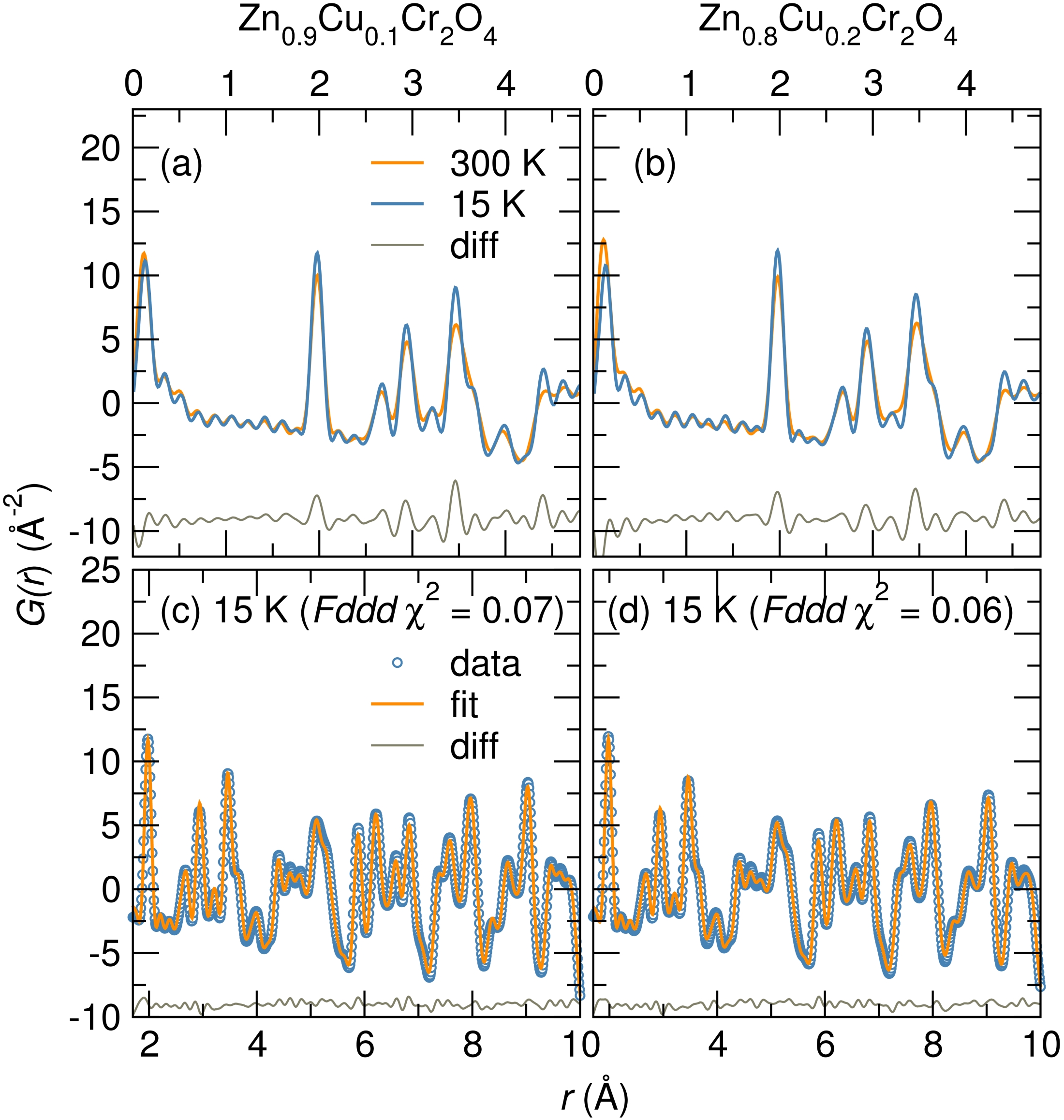}\\
\caption{(Color online) Pair distribution functions for \zcuco\,(a) and \zcucuco\,(b) measured at 300\,K and 15\,K show that the local structure varies slightly from room temperature to low temperature. The low $r$ region is shown in (a) and (b) to point out the slight increase in intensity of the pair distribution function at low temperature. The difference between the 15\,K and the 300\,K pair distribution functions is shown at the bottom. Least squares refinement of the pair distribution functions of these compounds at 15\,K are well modeled by the $Fddd$ structure as shown in (c) and (d).} 
\label{fig:pdflt}
\end{figure} 

Geometric spin frustration drives a lattice distortion at the antiferromagnetic
ordering temperature of \zco\, while Jahn-Teller distortion of tetrahedral Cu$^{2+}$
in Zn$_{1-x}$Cu$_x$Cr$_2$O$_4$ drive structural distortions in \zcuco\, and
\zcucuco\, at 45\,K and 110\,K respectively (middle panel of Fig.
\ref{fig:zcucovt}). The lattice distortions of \zcuco\, and \zcucuco\, are shown by
the divergence of the coincident (511) and (333) reflections at the respective
distortion temperatures of these materials (Fig. \ref{fig:zcucovt}). The structural
changes of \zcuco\, and \zcucuco\, are decoupled from antiferromagnetic ordering
(Table \ref{tab:magzcuco}), nonetheless there is a change in slope of the inverse
susceptibility of these systems at the structural distortion temperatures.
Jahn-Teller distortion is enhanced in Zn$_{1-x}$Cu$_x$Cr$_2$O$_4$ with increase in
Cu$^{2+}$ content, occurring at higher temperatures and involving larger lattice
distortions (Fig. \ref{fig:zcucovt}). Like in Mg$_{1-x}$Cu$_x$Cr$_2$O$_4$, the
Jahn-Teller phases of \zcuco\,[Fig. \ref{fig:zcucolt} (a)] and \zcucuco\,[Fig.
\ref{fig:zcucolt} (d)] are well modeled by the orthorhombic $Fddd$ space group. The
complete structural descriptions of these compounds at room temperature and near
6\,K are tabulated in Table \ref{tab:rietveldzcuco} of the appendix. 

The large heat capacity anomaly of \zco\, at the spin-Jahn-Teller distortion
temperature [Fig. \ref{fig:zcucovt} (d)] evolves into a broad transition in
\zcuco\,[Fig. \ref{fig:zcucovt} (e)] and \zcucuco\,[Fig. \ref{fig:zcucovt} (f)] over
the temperature range were structural and magnetic changes take place.

\begin{figure}[h!]
\centering\includegraphics[width=8.5cm]{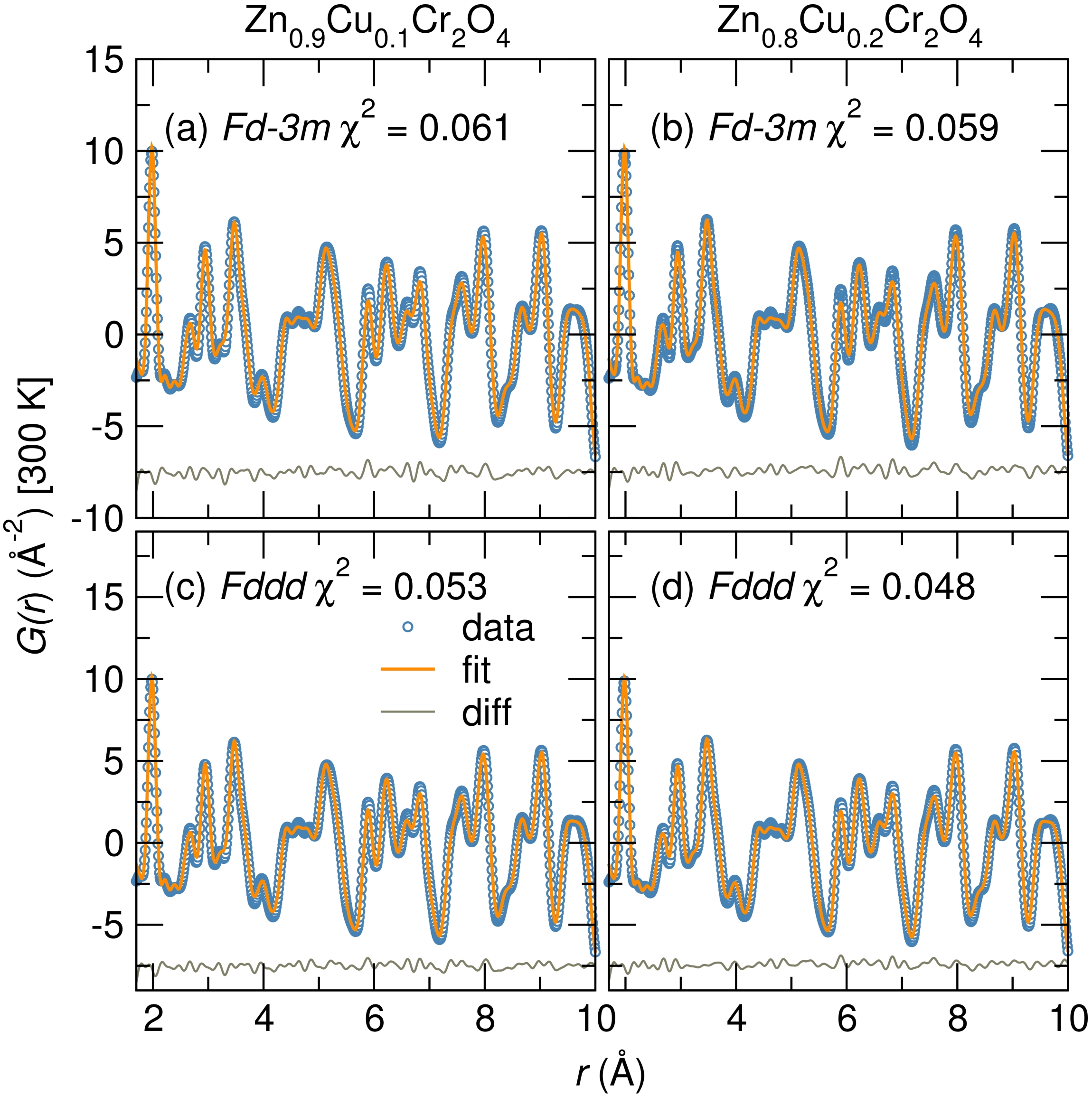}\\
\caption{(Color online) Least squares refinement of the pair distribution function of \zcuco\, and \zcucuco\, collected at 300\,K are indexed to the cubic $Fd\overline{3}m$ structure[(a) and (b)] and to the orthorhombic $Fddd$ structure[(c) and (d)]. A smaller difference curve and slightly better $\chi^2$ parameters are obtained when the orthorhombic model is applied to the room temperature data suggesting that dynamic Jahn-Teller distortion may be present at ambient temperature that becomes static at the Jahn-Teller distortion temperatures of these compounds.} 
\label{fig:pdfrt}
\end{figure}

\begin{figure}[h!]
\centering\includegraphics[width=8cm]{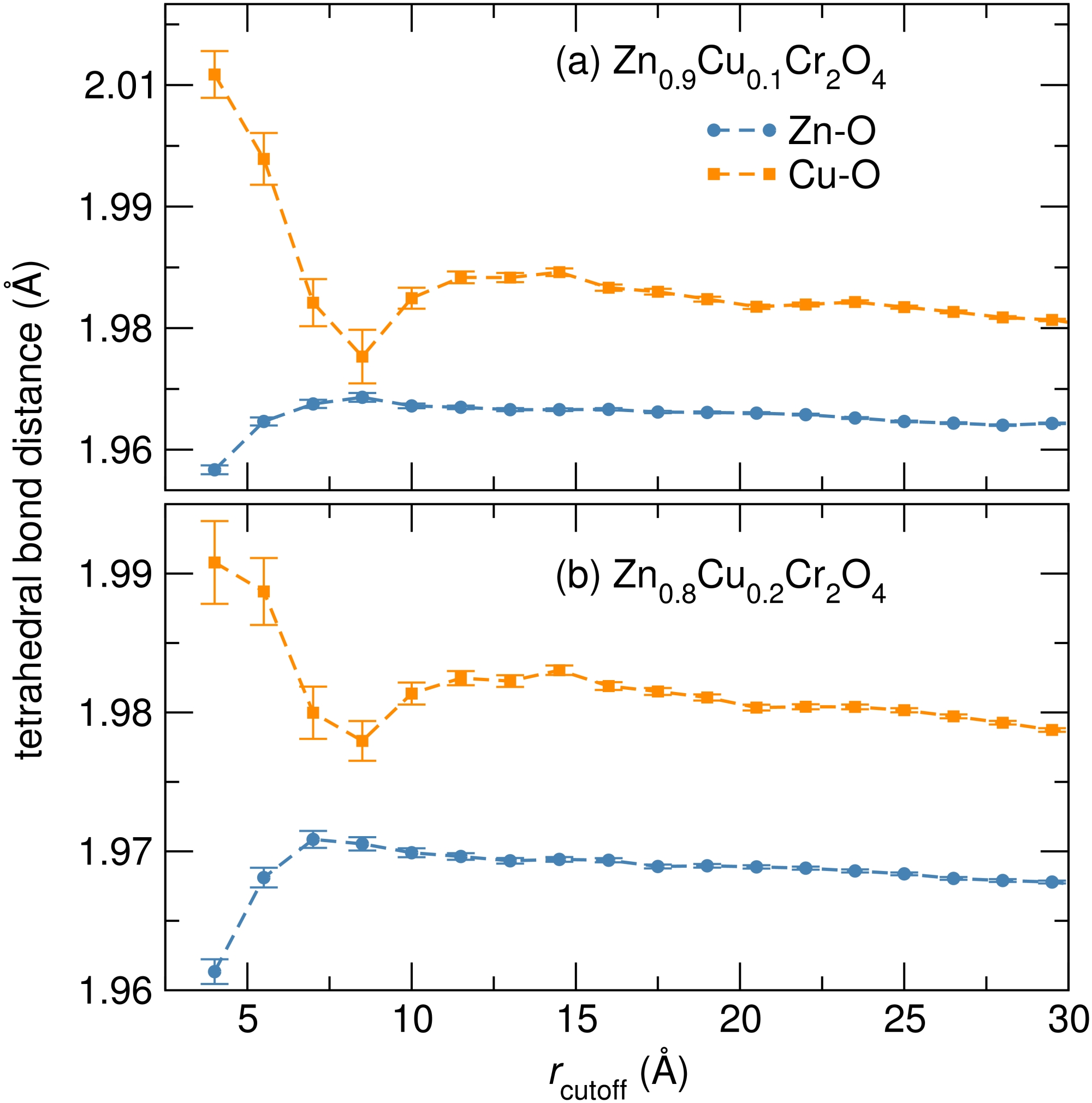}\\
\caption{(Color online) Least squares refinement of the pair distribution function of \zcuco\,(a) and \zcucuco\,(b) to a two phase model of \zco\, and \cuco\, shows that the environment around Zn$^{2+}$ varies from that around Cu$^{2+}$ at low $r$ as shown by the different Zn-O and Cu-O bond lengths of the ZnO$_4$ and CuO$_4$ tetrahedra. The phase fractions of the \zco\, model are 90$\%$ for \zcuco\, and 80$\%$ for \zcucuco. The differences in Zn-O and Cu-O bond lengths of the $A$O$_4$ tetrahedra are smaller at high $r$. } 
\label{fig:r}
\end{figure}

\begin{figure}[ht]
\centering\includegraphics[width=5.5cm]{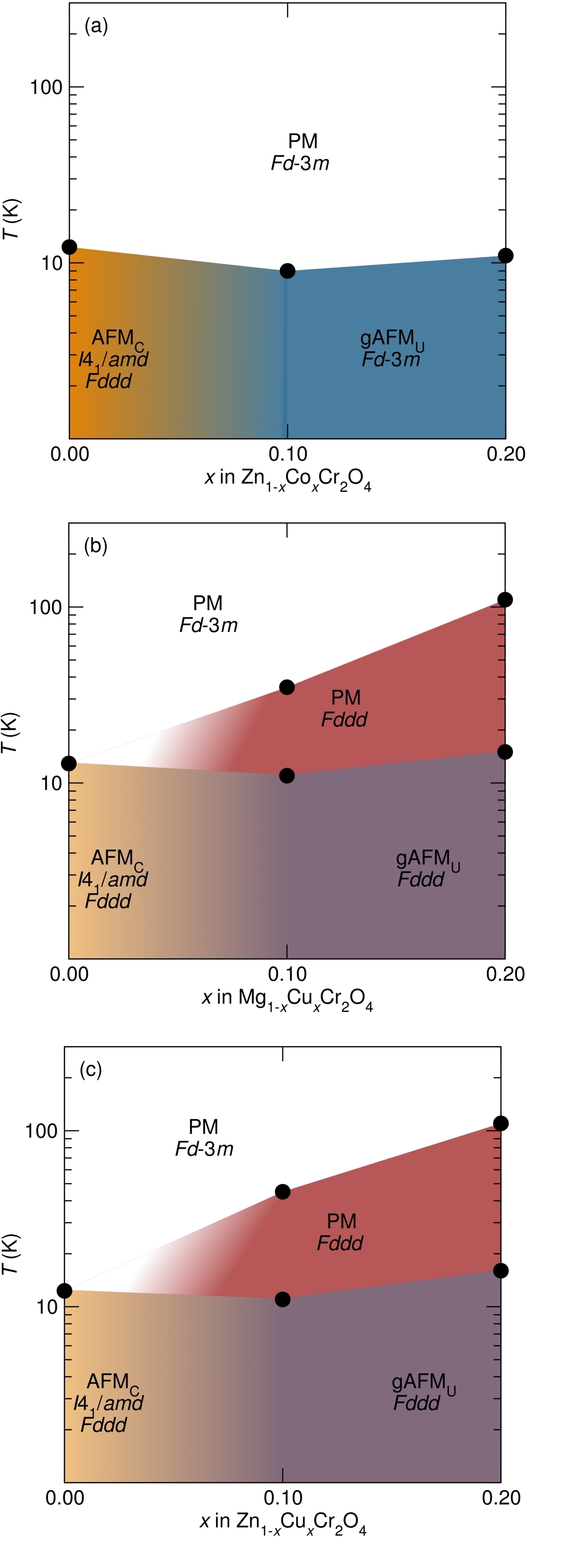}\\
\caption{(Color online) Temperature-composition phase diagrams of the spinel solid solutions Zn$_{1-x}$Co$_x$Cr$_2$O$_4$ (a), Mg$_{1-x}$Cu$_x$Cr$_2$O$_4$ (b), and Zn$_{1-x}$Cu$_x$Cr$_2$O$_4$ (c). At high temperatures all systems are paramagnetic(PM) and cubic in the space group $Fd\overline{3}m$. Magnetism evolves from frustrated compensated antiferromagnetism (AFM$_C$) in \zco\, and \mco\, to glassy uncompensated antiferromagnetism (gAFM$_U$) when $x$\,=0.2. A transition from cubic $Fd\overline{3}m$ to orthorhombic $Fddd$ symmetry occurs in Mg$_{1-x}$Cu$_x$Cr$_2$O$_4$  and Zn$_{1-x}$Cu$_x$Cr$_2$O$_4$  when $x$\,$\geq$\,0.1 due to Jahn-Teller distortions of tetrahedral CuO$_4$. We have recently reported that tetragonal $I4_1/amd$ and orthorhombic $Fddd$ structures coexist in the spin-Jahn-Teller phases of \mco\, and \zco.\cite{kemei_2013} } 
\label{fig:phase}
\end{figure}

We further explore the structural distortions of \zcucuco\, by performing sequential
Rietveld refinements from the high temperature cubic phase to the low temperature
orthorhombic phase. The cubic lattice constant diverges into three independent
orthorhombic lattice parameters at 110\,K [Fig. \ref{fig:lattice} (a)]. The $a$
lattice constant decreases steeply with temperature while the $b$ and $c$ lattice
constants increase. The orthorhombic distortion increases with decrease in
temperature. Although the structural distortions of Mg$_{1-x}$Cu$_x$Cr$_2$O$_4$ and
Zn$_{1-x}$Cu$_x$Cr$_2$O$_4$ occur primarily due to the Jahn-Teller activity of
tetrahedral Cu$^{2+}$, there is a distinct difference in the distortions observed in
the spinel solid solutions compared to the spinel CuCr$_2$O$_4$. Jahn-Teller
distortion in CuCr$_2$O$_4$ occurs near 853\,K and involve a cubic $Fd\overline{3}m$
to tetragonal $I4_1/amd$ lattice distortion.\cite{ye_1994}  Magnetostructural
coupling drives further structural distortion in CuCr$_2$O$_4$ from tetragonal
$I4_1/amd$ to orthorhombic $Fddd$ symmetry.\cite{suchomel_2012} In the solid
solutions Mg$_{1-x}$Cu$_x$Cr$_2$O$_4$ and Zn$_{1-x}$Cu$_x$Cr$_2$O$_4$, we observe a
cubic $Fd\overline{3}m$ to orthorhombic $Fddd$ distortion, completely bypassing the
tetragonal $I4_1/amd$ structure observed in CuCr$_2$O$_4$ and the lattice
distortions occur without the onset of magnetic ordering. The different character of
distortion in Mg$_{1-x}$Cu$_x$Cr$_2$O$_4$ and Zn$_{1-x}$Cu$_x$Cr$_2$O$_4$ compared
to \cuco\, is attributed to poor connectivity between CuO$_4$ tetrahedra. Dilute
randomly distributed CuO$_4$ tetrahedra in the solid solutions results in average
distortions in all axes of the unit cell and hence these systems adopt orthorhombic
symmetry in the Jahn-Teller phases. Group-subgroup relations show that structural
distortion from cubic $Fd\overline{3}m$ to orthorhombic $Fddd$ symmetry goes through
an intermediate tetragonal $I$4$_1$/$amd$ space group. It is plausible that the
Jahn-Teller ordering systems Mg$_{1-x}$Cu$_x$Cr$_2$O$_4$ and
Zn$_{1-x}$Cu$_x$Cr$_2$O$_4$ quickly go through the tetragonal $I$4$_1$/$amd$
structure before adopting the orthorhombic structure. A similar cubic
$Fd\overline{3}m$ to orthorhombic $Fddd$ lattice distortion driven by charge
ordering has been observed in the cathode spinel material LiMn$_2$O$_4$ near room
temperature.\cite{rodriguez_1998} 

The \zcucuco\, unit cell contracts with decrease in temperature as reflected in Fig.
\ref{fig:lattice} (b). There is a slight change in slope of the unit cell volume at
the structural distortion temperature. This small change in slope of the cell volume
at 110\,K and the broad heat capacity anomaly of \zcucuco\, suggest it undergoes a
second-order structural distortion.

The $A$O$_4$ tetrahedra of the cubic phase of \zcucuco\, are distorted in the
orthorhombic phase. Specifically, a single $A$-O distance is preserved while two
distinct O-$A$-O bond angles of 111.898$^{\circ}$ and 105.99$^{\circ}$ emerge from
the ideal tetrahedral angle 109.47$^{\circ}$ of the cubic phase. The overall effect
of these angle distortions in \zcucuco\, is a compression of the tetrahedra. This is
illustrated in Fig. \ref{fig:vesta} (a) where the ideal $A$O$_4$ tetrahedra of the
cubic phase leave small voids in the surrounding CrO$_6$ network while the flattened
$A$O$_4$ tetrahedra of the orthorhombic phase completely fill the tetrahedral voids
[Fig. \ref{fig:vesta}(b)]. The compression of $A$O$_4$ tetrahedra is similar to the
flattening of CuO$_4$ tetrahedra in the orthorhombic phase of
CuCr$_2$O$_4$.\cite{suchomel_2012} The $A$O$_4$ distortions of \zcucuco\, distort
the surrounding CrO$_6$ matrix. Three distinct Cr-O bond distances and O-Cr-O bond
angles emerge in the orthorhombic phase of \zcucuco\, compared to the cubic phase
where there are no bond length or bond angle distortions. The distortions of the Cr
sublattice in  Mg$_{1-x}$Cu$_x$Cr$_2$O$_4$ and Zn$_{1-x}$Cu$_x$Cr$_2$O$_4$ do not
lift spin degeneracy and magnetic ordering in these materials still takes place
below 20\,K (Table \ref{tab:magmcuco} and \ref{tab:magzcuco}).

Real space structural descriptions of Zn$_{1-x}$Cu$_x$Cr$_2$O$_4$ for $x$\,$>$\,0
give insights as to the nature of the Jahn-Teller distortions in these compounds.
There are differences in the pair distribution functions of \zcuco\, and \zcucuco\,
collected at room temperature and at 15\,K as shown in Fig. \ref{fig:pdflt} (a) and
(b). The differences are mainly in the intensity of atom pair correlations; at low
temperature, the distribution functions have slightly higher intensity than at room
temperature where atomic vibrations broaden the pair distribution function [Fig.
\ref{fig:pdflt} (a) and (b)]. The low temperature average structural model,
orthorhombic $Fddd$, describes the local structure of \zcuco\, and \zcucuco\, at
15\,K [Fig. \ref{fig:pdflt} (c) and (d)]. 

Combined average and local structure studies can distinguish whether Jahn-Teller
distortions occur spontaneously at the average structure distortion temperatures of
\zcuco\, and \zcucuco\, or whether local distortions of CuO$_4$ tetrahedra persist
in the cubic phases of these materials with these distortions becoming cooperative
at the Jahn-Teller distortion temperature. In Figure \ref{fig:pdfrt} we model the
room temperature pair distribution functions of \zcuco\,(a) and \zcucuco\,(b) to the
cubic average structure model $Fd\overline{3}m$ and to the Jahn-Teller distorted
orthorhombic $Fddd$ structure.  At room temperature, the cubic $Fd\overline{3}m$ fit
yields slightly larger goodness-of-fit parameters compared to the lower symmetry
$Fddd$ fits [Fig. \ref{fig:pdfrt} (c) and (d)]. The better description of the local
structure of these compounds at room temperature by the lower symmetry structural
model suggests that local CuO$_4$ distortions are present in the cubic phases of
\zcuco\, and \zcucuco\, and that these distortions become cooperative at the
respective Jahn-Teller distortion temperatures of these systems. The presence of
local distortions in \zcuco\, and \zcucuco\, is further corroborated by least
squares refinements of the pair distribution functions to a structural model of two
cubic phases assigned to either \zco\, or \cuco [Fig. \ref{fig:r} (a) and (b)].
Structural models consisting of stoichiometrically weighted end-member structures
have been previously successfully employed to describe the pair distribution
function of the frustrated spinel CoAl$_{1.6}$Ga$_{0.4}$O$_4$ at low
$r$.\cite{Melot_2009b} In the two phase refinement, the \zco\, and \cuco\,
structural models are scaled to correlate with the mole fractions of Zn$^{2+}$ and
Cu$^{2+}$ and only the lattice parameters and Zn$^{2+}$/Cu$^{2+}$ thermal parameters
are allowed to vary. There is a large difference in the Zn-O and Cu-O bond lengths
in the two phases at low $r_{\text{cutoff}}$ for both \zcuco\, and \zcucuco\, and
this difference decreases at high $r_{\text{cutoff}}$ (Fig. \ref{fig:r}). This
suggests that there are local distortions in \zcuco\, and \zcucuco\, that are best
modeled by differentiating the environment around Zn$^{2+}$ and Cu$^{2+}$. As one
examines the pair distribution functions to higher $r_{\text{cutoff}}$, these local
distortions are averaged out becoming less apparent as observed by the smaller
difference in $A$-O bond lengths of \zco\, and \cuco. The presence of local
distortions in \zcuco\, and \zcucuco\, is in good agreement with total scattering
studies of the spinels Mg$_{1-x}$Cu$_x$Cr$_2$O$_4$ by Shoemaker and Seshadri that
show more distortions of the local CuO$_4$ environments compared to MgO$_4$
environments in this spinel solid solution.\cite{shoemaker_2010} 

Cu$^{2+}$ substitution on the non-magnetic $A$ sites of \mco\, and \zco\, has very
similar effects on the structure and magnetic properties of the resulting solid
solutions. Average structure distortions due to Jahn-Teller ordering of CuO$_4$
tetrahedra occur in Mg$_{1-x}$Cu$_x$Cr$_2$O$_4$ and Zn$_{1-x}$Cu$_x$Cr$_2$O$_4$
without accompanying long-range magnetic ordering. The distortions of CuO$_4$
tetrahedra are present in the cubic phases of Zn$_{1-x}$Cu$_x$Cr$_2$O$_4$ when
$x$\,$\geq$\,0.1 and become cooperative at the Jahn-Teller distortion temperature
where average structure distortions are observed. The Jahn-Teller distorted phases
are described by the orthorhombic $Fddd$ space group. The degeneracy of Cr-Cr bond
distances is broken when average structure distortions occur in
Mg$_{1-x}$Cu$_x$Cr$_2$O$_4$ and Zn$_{1-x}$Cu$_x$Cr$_2$O$_4$, however,
antiferromagnetic interactions remain largely frustrated with magnetic ordering
occurring below 20\,K.  We contrast the propensity for Jahn-Teller distortions
compared with spin-Jahn-Teller distortion in the systems 
Mg$_{1-x}$Cu$_{x}$Cr$_2$O$_4$ and Zn$_{1-x}$Cu$_{x}$Cr$_2$O$_4$ where only few
Cu$^{2+}$ cations drive Jahn-Teller distortions while these small concentrations of
magnetic ions on the non-magnetic $A$ sites of these materials completely suppress
spin-Jahn-Teller distortion. The evolution of structure and magnetism in the solid
solutions Zn$_{1-x}$Co$_x$Cr$_2$O$_4$, Mg$_{1-x}$Cu$_x$Cr$_2$O$_4$, and
Zn$_{1-x}$Cu$_x$Cr$_2$O$_4$ are summarized in the phase diagrams presented in Fig.
\ref{fig:phase}.

\subsection{Conclusions}
We report the effect of magnetic $A$ site substitutions on spin and structural 
ordering in MgCr$_2$O$_4$ and ZnCr$_2$O$_4$. We contrast the
effect of Co$^{2+}$ $3d^7$ substitutions in Zn$_{1-x}$Co$_x$Cr$_2$O$_4$ with
Cu$^{2+}$ $3d^9$ substitutions in Mg$_{1-x}$Cu$_x$Cr$_2$O$_4$ and
Zn$_{1-x}$Cu$_x$Cr$_2$O$_4$. Spin disorder induced by Co$^{2+}$ $3d^7$ substitutions
in Zn$_{1-x}$Co$_x$Cr$_2$O$_4$ suppress the spin-Jahn-Teller distortion of
ZnCr$_2$O$_4$. On the other hand, spin and lattice disorder due to Cu$^{2+}$ $3d^9$
substitutions in Mg$_{1-x}$Cu$_x$Cr$_2$O$_4$ and Zn$_{1-x}$Cu$_x$Cr$_2$O$_4$ induce
Jahn-Teller distortions in the paramagnetic phases of these compounds yet
antiferromagnetic interactions in these systems remain frustrated with long-range
magnetic ordering occuring below 20\,K with no accompanying structural
transformations. In other words, the Jahn-Teller active Cu$^{2+}$ ions decouple 
structural and magnetic ordering, even when only substituted in small amounts.
The low-temperature nuclear structure of Cu$^{2+}$ substituted
MgCr$_2$O$_4$ and ZnCr$_2$O$_4$ is orthorhombic $Fddd$. Analysis of distortions in
\zcucuco\, indicate a flattening of $A$O$_4$ tetrahedra in the orthorhombic phase.
Total neutron scattering studies of Zn$_{1-x}$Cu$_x$Cr$_2$O$_4$ suggest that
$A$O$_4$ are likely distorted locally at room temperature with these distortions
becoming cooperative where average structure distortions are observed. Addition of
magnetic Co$^{2+}$ and Cu$^{2+}$ induce uncompensated antiferromagnetic interactions
in Zn$_{1-x}$Co$_x$Cr$_2$O$_4$, Mg$_{1-x}$Cu$_x$Cr$_2$O$_4$, and
Zn$_{1-x}$Cu$_x$Cr$_2$O$_4$. Compounds with dilute $A$ site spins have broad heat
capacity features suggesting remanent disorder in these materials. We find that
spin-Jahn-Teller ordering is extremely sensitive to spin disorder while Jahn-Teller
ordering is robust, and occurs even when only few Jahn-Teller active cations are
substituted into the spinel structure.

\subsection{Acknowledgements} 
This project was supported by the NSF through the DMR 1105301. MCK is supported by
the Schlumberger Foundation Faculty for the Future fellowship. We acknowledge the
use of shared experimental facilities of the Materials Research Laboratory: an NSF
MRSEC, supported by NSF DMR 1121053. The 11-BM beamline at the Advanced Photon
Source is supported by the DOE, Office of Science, Office of Basic Energy Sciences,
under Contract No. DE-AC0206CH11357. This work benefited from the use of NPDF at the
Los Alamos Neutron Scattering Center at Los Alamos National Laboratory, funded by
DOE Office of Basic Energy Sciences; LANL is operated by Los Alamos National
Security LLC under DE-AC52-06NA25396. MCK acknowledges helpful discussions with
Jason E. Douglas and Brent C. Melot.

\appendix*

\begin{table*}
\caption{\label{tab:rietveldZnCoCr2O4} Structural parameters of \zcoco\, and \zcococo\, obtained from Rietveld refinement of high-resolution synchrotron powder X-ray diffraction collected at 295\,K and near 6\,K and modeled to the cubic space group $Fd\overline{3}m$.}
\centering
\begin{tabular}{lllllllllll }
\hline
&& \zcoco\/ & & & & \zcococo\/ &  \ \\
\hline
\hline
Temperature (K) 				&  & 5.5 & 295 & & & 5.6 & 295 \ \\
Setting 						&	& origin 2 & origin 2 & & & origin 2 & origin 2\ \\
$Z$  							& & 8 & 8 & & & 8 & 8\ \\
$a$(\AA) 						& &8.32037(2)&8.327641(1)& & & 8.32038(1) & 8.327868(5)\ \\
Vol/ (\AA$^3$)					&  & 576.008(4) & 577.519(3) & & & 576.009(2) & 577.566(1)\ \\
Zn/Co 							& & $8a$ (1/8,\,1/8\,1/8) & $8a$ (1/8,\,1/8,\,1/8))& & & $8a$ (1/8,\,1/8\,1/8)& $8a$ (1/8,\,1/8,\,1/8)\ \\
$U_{iso}$ ($10^2$ \AA$^2$) 		& & 0.24(1) & 0.389(5) & & & 0.128(7)& 0.398(5)\ \\
Cr & & $16d$ (1/2,\,1/2, \,1/2) & $16d$ (1/2,\,1/2,\,1/2)& & & $16d$ (1/2,\,1/2, \,1/2) & $16d$ (1/2,\,1/2,\,1/2) \ \\
$U_{iso}$ ($10^2$ \AA$^2$) 		& & 0.196(9) & 0.213(4)& & & 0.145(6) & 0.230(4)\ \\
O 								& & $32h$ ($x,y,z$) & $32e$ ($x,x,x$)& & & $32h$ ($x,y,z$) & $32e$ ($x,x,x$)\ \\
 								& & $x$ 0.2612(1)& $x$ 0.26205(4)& & & 0.26208(7) & 0.26221(5)\ \\
 
$U_{iso}$ ($10^2$ \AA$^2$) 		& & 0.54(3) & 0.32(1)& & & 0.35(2)& 0.360(1)\ \\

$\chi^2$ 						& & 15.73 & 5.263& & &6.976& 7.434\ \\
$R_p$($\%$)						& & 8.10 & 9.12& & & 7.54& 9.33 \\
$R_{wp}$($\%$)					& & 10.12 & 11.54 & & & 9.43& 12.56\ \\
\hline
\hline
\end{tabular}
\end{table*}
\begin{table*}
\caption{\label{tab:rietveldmcco} Structural parameters of \mcuco\, and \mcucuco\, obtained from Rietveld refinement of high-resolution synchrotron powder X-ray diffraction collected at 295\,K and near 6\,K.}
\centering
\begin{tabular}{lllllllllll }
\hline
&& \mcuco\/ & & & & \mcucuco\/ &  \ \\
\hline
&&Orthorhombic & Cubic & & & Orthorhombic & Cubic \	  \\
\hline
\hline
Temperature (K) 				&  & 6.5 & 295 & & & 5.7 & 295 \ \\
Space group 					& & $Fddd$& $Fd\overline{3}m$& & & $Fddd$ & $Fd\overline{3}m$\ \\
Setting 						&	& origin 2 & origin 2 & & & origin 2 & origin 2\ \\
$Z$  							& & 8 & 8 & & & 8 & 8\ \\
$a$(\AA) 						& &8.293741(1)&8.332613(3)& & & 8.231110(8) & 8.330362(3)\ \\
$b$(\AA) 						& &8.335834(1)&8.332613(3)& & & 8.360256(9) & 8.330362(3)\ \\
$c$(\AA) 						& &8.3488(5)&8.332613(3)& & & 8.373391(5)& 8.330362(3)\ \\
Vol/ (\AA$^3$)					&  & 577.196(4) & 578.554(1) & & & 576.208(4) & 578.085(1)\ \\
Mg/Cu 							& & $8a$ (1/8,\,1/8\,1/8) & $8a$ (1/8,\,1/8,\,1/8))& & & $8a$ (1/8,\,1/8\,1/8)& $8a$ (1/8,\,1/8,\,1/8)\ \\
$U_{iso}$ ($10^2$ \AA$^2$) 		& & 0.384(2) & 0.651(7)& & & 0.371(2)& 0.696(7)\ \\
Cr & & $16d$ (1/2,\,1/2, \,1/2) & $16d$ (1/2,\,1/2,\,1/2)& & & $16d$ (1/2,\,1/2, \,1/2) & $16d$ (1/2,\,1/2,\,1/2) \ \\
$U_{iso}$ ($10^2$ \AA$^2$) 		& & 0.17(9) & 0.321(3)& & & 0.102(1) & 0.344(3)\ \\
O 								& & $32h$ ($x,y,z$) & $32e$ ($x,x,x$)& & & $32h$ ($x,y,z$) & $32e$ ($x,x,x$)\ \\
 								& & $x$ 0.26284(2)& $x$ 0.26161(3)& & & 0.260090(1) & 0.26170(4)\ \\
 								& & $y$ 0.25845(3) & $y$ 0.26161(3)& & & 0.260627(2) & 0.26170(4)\ \\
 								&  &$z$ 0.262891(2) & $z$ 0.26161(3)& & & 0.263825(2) & 0.26170(4)\ \\
$U_{iso}$ ($10^2$ \AA$^2$) 		& & 0.144(2) & 0.651(7)& & & 0.25(0)& 0.59(1)\ \\
Cr$_2$O$_3$ wt$\%$ 				& & 0.89(8) & 0.89(8) & & & 0.87(7) & 0.87(7) \ \\
$\chi^2$ 						& & 4.37 & 2.550& & &6.81& 2.571\ \\
$R_p$($\%$)						& & 6.7 & 5.98& & & 6.29& 6.23\ \\
$R_{wp}$($\%$)					& & 9.16 & 8.12 & & & 8.14& 8.15\ \\
\hline
\hline
\end{tabular}
\end{table*}
\newpage
\begin{table*}
\caption{\label{tab:rietveldzcuco} Structural parameters of \zcuco\, and \zcucuco\, obtained from Rietveld refinement of high-resolution synchrotron powder X-ray diffraction collected at 295\,K and near 6\,K.}
\centering
\begin{tabular}{lllllllllll }
\hline
&& \zcuco\/ & & & & \zcucuco\/ &  \ \\
\hline
&&Orthorhombic & Cubic & & & Orthorhombic & Cubic \	  \\
\hline
\hline
Temperature (K) 				&  & 5.6 & 295 & & & 5.9 & 295 \ \\
Space group 					& & $Fddd$& $Fd\overline{3}m$& & & $Fddd$ & $Fd\overline{3}m$\ \\
Setting 						&	& origin 2 & origin 2 & & & origin 2 & origin 2\ \\
$Z$  							& & 8 & 8 & & & 8 & 8\ \\
$a$(\AA) 						& &8.328150(7)&8.326101(6)& & & 8.244895(8) & 8.324244(7)\ \\
$b$(\AA) 						& &8.335862(5)&8.326101(6)& & & 8.358441(5) & 8.324244(7)\ \\
$c$(\AA) 						& &8.289697(5)&8.326101(6)& & & 8.3409(1)& 8.324244(7)\ \\
Vol/ (\AA$^3$)					&  & 575.490(3) & 577.198(1) & & & 574.809(5) & 576.812(2)\ \\
Zn/Cu 							& & $8a$ (1/8,\,1/8\,1/8) & $8a$ (1/8,\,1/8,\,1/8))& & & $8a$ (1/8,\,1/8\,1/8)& $8a$ (1/8,\,1/8,\,1/8)\ \\
$U_{iso}$ ($10^2$ \AA$^2$) 		& & 0.309(8) & 0.455(5)& & & 0.281(13)& 0.470(5)\ \\
Cr & & $16d$ (1/2,\,1/2, \,1/2) & $16d$ (1/2,\,1/2,\,1/2)& & & $16d$ (1/2,\,1/2, \,1/2) & $16d$ (1/2,\,1/2,\,1/2) \ \\
$U_{iso}$ ($10^2$ \AA$^2$) 		& & 0.297(8) & 0.259(4)& & & 0.254(13) & 0.305(5)\ \\
O 								& & $32h$ ($x,y,z$) & $32e$ ($x,x,x$)& & & $32h$ ($x,y,z$) & $32e$ ($x,y,z$)\ \\
 								& & $x$ 0.2577(3)& $x$ 0.262015(47)& & & 0.260315(184) & 0.261641(53)\ \\
 								& & $y$ 0.2642(3) & $y$ 0.262015(47)& & & 0.266652(262) & 0.261641(53)\ \\
 								&  &$z$ 0.2628(2) & $z$ 0.262015(47)& & & 0.257599(340) & 0.261641(53))\ \\
$U_{iso}$ ($10^2$ \AA$^2$) 		& & 0.329(21) & 0.411(14)& & & 0.304(33)& 0.527(16)\ \\
Cr$_2$O$_3$ wt$\%$ 				& & 0.97(2) & 0.97(2) & & & 1.05(15) & 1.05(15) \ \\
$\chi^2$ 						& & 4.494 & 6.582& & &7.808& 7.805\ \\
$R_p$($\%$)						& & 2.52 & 8.53& & & 3.81& 7.20\ \\
$R_{wp}$($\%$)					& & 3.49 & 11.97 & & & 5.85& 11.30\ \\
\hline
\hline
\end{tabular}
\end{table*}

\clearpage
\bibliography{Spinels1_v5}

\begin{thebibliography}{37}
\expandafter\ifx\csname natexlab\endcsname\relax\def\natexlab#1{#1}\fi
\expandafter\ifx\csname bibnamefont\endcsname\relax
  \def\bibnamefont#1{#1}\fi
\expandafter\ifx\csname bibfnamefont\endcsname\relax
  \def\bibfnamefont#1{#1}\fi
\expandafter\ifx\csname citenamefont\endcsname\relax
  \def\citenamefont#1{#1}\fi
\expandafter\ifx\csname url\endcsname\relax
  \def\url#1{\texttt{#1}}\fi
\expandafter\ifx\csname urlprefix\endcsname\relax\def\urlprefix{URL }\fi
\providecommand{\bibinfo}[2]{#2}
\providecommand{\eprint}[2][]{\url{#2}}

\bibitem[{\citenamefont{Pascut et~al.}(2011)\citenamefont{Pascut, Coldea,
  Radaelli, Bombardi, Beutier, Mazin, Johannes, and Jansen}}]{pascut_2011}
\bibinfo{author}{\bibfnamefont{G.~L.} \bibnamefont{Pascut}},
  \bibinfo{author}{\bibfnamefont{R.}~\bibnamefont{Coldea}},
  \bibinfo{author}{\bibfnamefont{P.~G.} \bibnamefont{Radaelli}},
  \bibinfo{author}{\bibfnamefont{A.}~\bibnamefont{Bombardi}},
  \bibinfo{author}{\bibfnamefont{G.}~\bibnamefont{Beutier}},
  \bibinfo{author}{\bibfnamefont{I.~I.} \bibnamefont{Mazin}},
  \bibinfo{author}{\bibfnamefont{M.~D.} \bibnamefont{Johannes}},
  \bibnamefont{and} \bibinfo{author}{\bibfnamefont{M.}~\bibnamefont{Jansen}},
  \bibinfo{journal}{Phys. Rev. Lett} \textbf{\bibinfo{volume}{106}},
  \bibinfo{pages}{157206} (\bibinfo{year}{2011}).

\bibitem[{\citenamefont{Lee et~al.}(2002)\citenamefont{Lee, Broholm, Ratcliff,
  Gasparovic, Huang, Kim, and Cheong}}]{lee_2002}
\bibinfo{author}{\bibfnamefont{S.-H.} \bibnamefont{Lee}},
  \bibinfo{author}{\bibfnamefont{C.}~\bibnamefont{Broholm}},
  \bibinfo{author}{\bibfnamefont{W.}~\bibnamefont{Ratcliff}},
  \bibinfo{author}{\bibfnamefont{G.}~\bibnamefont{Gasparovic}},
  \bibinfo{author}{\bibfnamefont{Q.}~\bibnamefont{Huang}},
  \bibinfo{author}{\bibfnamefont{T.~H.} \bibnamefont{Kim}}, \bibnamefont{and}
  \bibinfo{author}{\bibfnamefont{S.-W.} \bibnamefont{Cheong}},
  \bibinfo{journal}{Nature} \textbf{\bibinfo{volume}{418}},
  \bibinfo{pages}{856} (\bibinfo{year}{2002}).

\bibitem[{\citenamefont{Ehrenberg et~al.}(2002)\citenamefont{Ehrenberg, Knapp,
  Baehtz, and Klemme}}]{ehrenberg_2002}
\bibinfo{author}{\bibfnamefont{H.}~\bibnamefont{Ehrenberg}},
  \bibinfo{author}{\bibfnamefont{M.}~\bibnamefont{Knapp}},
  \bibinfo{author}{\bibfnamefont{C.}~\bibnamefont{Baehtz}}, \bibnamefont{and}
  \bibinfo{author}{\bibfnamefont{S.}~\bibnamefont{Klemme}},
  \bibinfo{journal}{Powder Diffr.} \textbf{\bibinfo{volume}{17}},
  \bibinfo{pages}{230} (\bibinfo{year}{2002}).

\bibitem[{\citenamefont{Martin et~al.}(2008)\citenamefont{Martin, Williams,
  Gordon, Klemme, and Attfield}}]{martin_2008}
\bibinfo{author}{\bibfnamefont{L.-S.} \bibnamefont{Martin}},
  \bibinfo{author}{\bibfnamefont{A.~J.} \bibnamefont{Williams}},
  \bibinfo{author}{\bibfnamefont{C.~D.} \bibnamefont{Gordon}},
  \bibinfo{author}{\bibfnamefont{S.}~\bibnamefont{Klemme}}, \bibnamefont{and}
  \bibinfo{author}{\bibfnamefont{J.~P.} \bibnamefont{Attfield}},
  \bibinfo{journal}{J. Phys.: Condens. Matter} \textbf{\bibinfo{volume}{20}},
  \bibinfo{pages}{104238} (\bibinfo{year}{2008}).

\bibitem[{\citenamefont{Rovers et~al.}(2002)\citenamefont{Rovers, Kyriakou,
  Dabkowska, Luke, Larkin, and Savici}}]{rovers_2002}
\bibinfo{author}{\bibfnamefont{M.~T.} \bibnamefont{Rovers}},
  \bibinfo{author}{\bibfnamefont{P.~P.} \bibnamefont{Kyriakou}},
  \bibinfo{author}{\bibfnamefont{H.~A.} \bibnamefont{Dabkowska}},
  \bibinfo{author}{\bibfnamefont{G.~M.} \bibnamefont{Luke}},
  \bibinfo{author}{\bibfnamefont{M.~I.} \bibnamefont{Larkin}},
  \bibnamefont{and} \bibinfo{author}{\bibfnamefont{A.~T.}
  \bibnamefont{Savici}}, \bibinfo{journal}{Phys. Rev. B}
  \textbf{\bibinfo{volume}{66}}, \bibinfo{pages}{174434}
  (\bibinfo{year}{2002}).

\bibitem[{\citenamefont{Kemei et~al.}(2013)\citenamefont{Kemei, Barton,
  Moffitt, Gaultois, Kurzman, Seshadri, Suchomel, and Kim}}]{kemei_2013}
\bibinfo{author}{\bibfnamefont{M.~C.} \bibnamefont{Kemei}},
  \bibinfo{author}{\bibfnamefont{P.~T.} \bibnamefont{Barton}},
  \bibinfo{author}{\bibfnamefont{S.~L.} \bibnamefont{Moffitt}},
  \bibinfo{author}{\bibfnamefont{M.~W.} \bibnamefont{Gaultois}},
  \bibinfo{author}{\bibfnamefont{J.~A.} \bibnamefont{Kurzman}},
  \bibinfo{author}{\bibfnamefont{R.}~\bibnamefont{Seshadri}},
  \bibinfo{author}{\bibfnamefont{M.~R.} \bibnamefont{Suchomel}},
  \bibnamefont{and} \bibinfo{author}{\bibfnamefont{Y.}~\bibnamefont{Kim}},
  \bibinfo{journal}{J. Phys.: Condens. Matter} \textbf{\bibinfo{volume}{25}},
  \bibinfo{pages}{326001} (\bibinfo{year}{2013}).

\bibitem[{\citenamefont{Ramirez}(1994)}]{ramirez_1994}
\bibinfo{author}{\bibfnamefont{A.~P.} \bibnamefont{Ramirez}},
  \bibinfo{journal}{Annu. Rev. Mater. Sci.} \textbf{\bibinfo{volume}{24}},
  \bibinfo{pages}{453} (\bibinfo{year}{1994}).

\bibitem[{\citenamefont{Kagomiya et~al.}(2002)\citenamefont{Kagomiya, Sawa,
  Siratori, Kohn, Toki, Hata, and Kita}}]{kagomiya_2002}
\bibinfo{author}{\bibfnamefont{I.}~\bibnamefont{Kagomiya}},
  \bibinfo{author}{\bibfnamefont{H.}~\bibnamefont{Sawa}},
  \bibinfo{author}{\bibfnamefont{K.}~\bibnamefont{Siratori}},
  \bibinfo{author}{\bibfnamefont{K.}~\bibnamefont{Kohn}},
  \bibinfo{author}{\bibfnamefont{M.}~\bibnamefont{Toki}},
  \bibinfo{author}{\bibfnamefont{Y.}~\bibnamefont{Hata}}, \bibnamefont{and}
  \bibinfo{author}{\bibfnamefont{E.}~\bibnamefont{Kita}},
  \bibinfo{journal}{Ferroelectrics} \textbf{\bibinfo{volume}{268}},
  \bibinfo{pages}{327} (\bibinfo{year}{2002}).

\bibitem[{\citenamefont{Chung et~al.}(2005)\citenamefont{Chung, Matsuda, Lee,
  Kakurai, Ueda, Sato, Takagi, Hong, and Park}}]{chung_2005}
\bibinfo{author}{\bibfnamefont{J.-H.} \bibnamefont{Chung}},
  \bibinfo{author}{\bibfnamefont{M.}~\bibnamefont{Matsuda}},
  \bibinfo{author}{\bibfnamefont{S.-H.} \bibnamefont{Lee}},
  \bibinfo{author}{\bibfnamefont{K.}~\bibnamefont{Kakurai}},
  \bibinfo{author}{\bibfnamefont{H.}~\bibnamefont{Ueda}},
  \bibinfo{author}{\bibfnamefont{T.~J.} \bibnamefont{Sato}},
  \bibinfo{author}{\bibfnamefont{H.}~\bibnamefont{Takagi}},
  \bibinfo{author}{\bibfnamefont{K.~P.} \bibnamefont{Hong}}, \bibnamefont{and}
  \bibinfo{author}{\bibfnamefont{S.}~\bibnamefont{Park}},
  \bibinfo{journal}{Phys. Rev. Lett.} \textbf{\bibinfo{volume}{95}},
  \bibinfo{pages}{247204} (\bibinfo{year}{2005}).

\bibitem[{\citenamefont{Aguilar et~al.}(2008)\citenamefont{Aguilar, Sushkov,
  Choi, Cheong, and Drew}}]{aguilar_2008}
\bibinfo{author}{\bibfnamefont{R.~V.} \bibnamefont{Aguilar}},
  \bibinfo{author}{\bibfnamefont{A.~B.} \bibnamefont{Sushkov}},
  \bibinfo{author}{\bibfnamefont{Y.~J.} \bibnamefont{Choi}},
  \bibinfo{author}{\bibfnamefont{S.~W.} \bibnamefont{Cheong}},
  \bibnamefont{and} \bibinfo{author}{\bibfnamefont{H.~D.} \bibnamefont{Drew}},
  \bibinfo{journal}{Phys. Rev. B} \textbf{\bibinfo{volume}{77}},
  \bibinfo{pages}{092412} (\bibinfo{year}{2008}).

\bibitem[{\citenamefont{Ueda et~al.}(2006)\citenamefont{Ueda, Mitamura, Goto,
  and Ueda}}]{Ueda_2006}
\bibinfo{author}{\bibfnamefont{H.}~\bibnamefont{Ueda}},
  \bibinfo{author}{\bibfnamefont{H.}~\bibnamefont{Mitamura}},
  \bibinfo{author}{\bibfnamefont{T.}~\bibnamefont{Goto}}, \bibnamefont{and}
  \bibinfo{author}{\bibfnamefont{Y.}~\bibnamefont{Ueda}},
  \bibinfo{journal}{Phys. Rev. B} \textbf{\bibinfo{volume}{73}},
  \bibinfo{pages}{094415} (\bibinfo{year}{2006}).

\bibitem[{\citenamefont{Melot et~al.}(2009{\natexlab{a}})\citenamefont{Melot,
  Drewes, Seshadri, Stoudenmire, and Ramirez}}]{melot_2009}
\bibinfo{author}{\bibfnamefont{B.~C.} \bibnamefont{Melot}},
  \bibinfo{author}{\bibfnamefont{J.~E.} \bibnamefont{Drewes}},
  \bibinfo{author}{\bibfnamefont{R.}~\bibnamefont{Seshadri}},
  \bibinfo{author}{\bibfnamefont{E.~M.} \bibnamefont{Stoudenmire}},
  \bibnamefont{and} \bibinfo{author}{\bibfnamefont{A.~P.}
  \bibnamefont{Ramirez}}, \bibinfo{journal}{J. Phys.: Condens. Matter}
  \textbf{\bibinfo{volume}{21}}, \bibinfo{pages}{216007}
  (\bibinfo{year}{2009}{\natexlab{a}}).

\bibitem[{\citenamefont{Kemei et~al.}(2012)\citenamefont{Kemei, Moffitt,
  Shoemaker, and Seshadri}}]{kemei_2012}
\bibinfo{author}{\bibfnamefont{M.~C.} \bibnamefont{Kemei}},
  \bibinfo{author}{\bibfnamefont{S.~L.} \bibnamefont{Moffitt}},
  \bibinfo{author}{\bibfnamefont{D.~P.} \bibnamefont{Shoemaker}},
  \bibnamefont{and} \bibinfo{author}{\bibfnamefont{R.}~\bibnamefont{Seshadri}},
  \bibinfo{journal}{J. Phys.: Condens. Matter} \textbf{\bibinfo{volume}{24}},
  \bibinfo{pages}{046003} (\bibinfo{year}{2012}).

\bibitem[{\citenamefont{Yan et~al.}(2008)\citenamefont{Yan, Macia, Jiang, Shen,
  He, and Wang}}]{yan_2008}
\bibinfo{author}{\bibfnamefont{L.}~\bibnamefont{Yan}},
  \bibinfo{author}{\bibfnamefont{F.}~\bibnamefont{Macia}},
  \bibinfo{author}{\bibfnamefont{Z.}~\bibnamefont{Jiang}},
  \bibinfo{author}{\bibfnamefont{J.}~\bibnamefont{Shen}},
  \bibinfo{author}{\bibfnamefont{L.}~\bibnamefont{He}}, \bibnamefont{and}
  \bibinfo{author}{\bibfnamefont{F.}~\bibnamefont{Wang}}, \bibinfo{journal}{J.
  Phys.: Condens. Matter} \textbf{\bibinfo{volume}{20}},
  \bibinfo{pages}{255203} (\bibinfo{year}{2008}).

\bibitem[{\citenamefont{LaForge et~al.}(2013)\citenamefont{LaForge, Pulido,
  Cava, Chan, and Ramirez}}]{laforge_2013}
\bibinfo{author}{\bibfnamefont{A.~D.} \bibnamefont{LaForge}},
  \bibinfo{author}{\bibfnamefont{S.~H.} \bibnamefont{Pulido}},
  \bibinfo{author}{\bibfnamefont{R.~J.} \bibnamefont{Cava}},
  \bibinfo{author}{\bibfnamefont{B.~C.} \bibnamefont{Chan}}, \bibnamefont{and}
  \bibinfo{author}{\bibfnamefont{A.~P.} \bibnamefont{Ramirez}},
  \bibinfo{journal}{Phys. Rev. Lett.} \textbf{\bibinfo{volume}{110}},
  \bibinfo{pages}{017203} (\bibinfo{year}{2013}).

\bibitem[{\citenamefont{Dutton et~al.}(2011)\citenamefont{Dutton, Huang,
  Tchernyshyov, Broholm, and Cava}}]{dutton_2011}
\bibinfo{author}{\bibfnamefont{S.~E.} \bibnamefont{Dutton}},
  \bibinfo{author}{\bibfnamefont{Q.}~\bibnamefont{Huang}},
  \bibinfo{author}{\bibfnamefont{O.}~\bibnamefont{Tchernyshyov}},
  \bibinfo{author}{\bibfnamefont{C.~L.} \bibnamefont{Broholm}},
  \bibnamefont{and} \bibinfo{author}{\bibfnamefont{R.~J.} \bibnamefont{Cava}},
  \bibinfo{journal}{Phys. Rev. B} \textbf{\bibinfo{volume}{83}},
  \bibinfo{pages}{064407} (\bibinfo{year}{2011}).

\bibitem[{\citenamefont{Shannon}(1976)}]{shannon_1976}
\bibinfo{author}{\bibfnamefont{R.~D.} \bibnamefont{Shannon}},
  \bibinfo{journal}{Acta Crystallogr., Sect. A: Found. Crystallogr.}
  \textbf{\bibinfo{volume}{32}}, \bibinfo{pages}{751} (\bibinfo{year}{1976}).

\bibitem[{\citenamefont{Suchomel et~al.}(2012)\citenamefont{Suchomel,
  Shoemaker, Ribaud, Kemei, and Seshadri}}]{suchomel_2012}
\bibinfo{author}{\bibfnamefont{M.~R.} \bibnamefont{Suchomel}},
  \bibinfo{author}{\bibfnamefont{D.~P.} \bibnamefont{Shoemaker}},
  \bibinfo{author}{\bibfnamefont{L.}~\bibnamefont{Ribaud}},
  \bibinfo{author}{\bibfnamefont{M.~C.} \bibnamefont{Kemei}}, \bibnamefont{and}
  \bibinfo{author}{\bibfnamefont{R.}~\bibnamefont{Seshadri}},
  \bibinfo{journal}{Phys. Rev. B} \textbf{\bibinfo{volume}{86}},
  \bibinfo{pages}{054406} (\bibinfo{year}{2012}).

\bibitem[{\citenamefont{Barton et~al.}(2013)\citenamefont{Barton, Seshadri,
  Llobet, and Suchomel}}]{barton_2013}
\bibinfo{author}{\bibfnamefont{P.~T.} \bibnamefont{Barton}},
  \bibinfo{author}{\bibfnamefont{R.}~\bibnamefont{Seshadri}},
  \bibinfo{author}{\bibfnamefont{A.}~\bibnamefont{Llobet}}, \bibnamefont{and}
  \bibinfo{author}{\bibfnamefont{M.~R.} \bibnamefont{Suchomel}},
  \bibinfo{journal}{Phys. Rev. B} \textbf{\bibinfo{volume}{88}},
  \bibinfo{pages}{024403} (\bibinfo{year}{2013}).

\bibitem[{\citenamefont{Shoemaker and Seshadri}(2010)}]{shoemaker_2010}
\bibinfo{author}{\bibfnamefont{D.~P.} \bibnamefont{Shoemaker}}
  \bibnamefont{and} \bibinfo{author}{\bibfnamefont{R.}~\bibnamefont{Seshadri}},
  \bibinfo{journal}{Phys. Rev. B} \textbf{\bibinfo{volume}{82}},
  \bibinfo{pages}{214107} (\bibinfo{year}{2010}).

\bibitem[{\citenamefont{Toby}(2001)}]{toby_expgui_2001}
\bibinfo{author}{\bibfnamefont{B.~H.} \bibnamefont{Toby}}, \bibinfo{journal}{J.
  Appl. Crystallogr.} \textbf{\bibinfo{volume}{34}}, \bibinfo{pages}{210}
  (\bibinfo{year}{2001}).

\bibitem[{\citenamefont{Larson and Dreele}(2000)}]{larson_2000}
\bibinfo{author}{\bibfnamefont{A.~C.} \bibnamefont{Larson}} \bibnamefont{and}
  \bibinfo{author}{\bibfnamefont{R.~B.~V.} \bibnamefont{Dreele}},
  \bibinfo{journal}{Los Alamos National Laboratory Report} pp.
  \bibinfo{pages}{86--748} (\bibinfo{year}{2000}).

\bibitem[{\citenamefont{Momma and Izumi}(2008)}]{momma_vesta_2008}
\bibinfo{author}{\bibfnamefont{K.}~\bibnamefont{Momma}} \bibnamefont{and}
  \bibinfo{author}{\bibfnamefont{F.}~\bibnamefont{Izumi}}, \bibinfo{journal}{J.
  Appl. Crystallogr.} \textbf{\bibinfo{volume}{41}}, \bibinfo{pages}{653}
  (\bibinfo{year}{2008}).

\bibitem[{\citenamefont{Peterson et~al.}(2000)\citenamefont{Peterson, Gutmann,
  Proffen, and Billinge}}]{peterson_2000}
\bibinfo{author}{\bibfnamefont{P.~F.} \bibnamefont{Peterson}},
  \bibinfo{author}{\bibfnamefont{M.}~\bibnamefont{Gutmann}},
  \bibinfo{author}{\bibfnamefont{T.}~\bibnamefont{Proffen}}, \bibnamefont{and}
  \bibinfo{author}{\bibfnamefont{S.~J.~L.} \bibnamefont{Billinge}},
  \bibinfo{journal}{J. Appl. Crystallogr.} \textbf{\bibinfo{volume}{33}},
  \bibinfo{pages}{1192} (\bibinfo{year}{2000}).

\bibitem[{\citenamefont{Farrow et~al.}(2007)\citenamefont{Farrow, Juhas, Liu,
  Bryndin, Bozin, Bloch, Proffen, and Billinge}}]{farrow_2007}
\bibinfo{author}{\bibfnamefont{C.~L.} \bibnamefont{Farrow}},
  \bibinfo{author}{\bibfnamefont{P.}~\bibnamefont{Juhas}},
  \bibinfo{author}{\bibfnamefont{J.~W.} \bibnamefont{Liu}},
  \bibinfo{author}{\bibfnamefont{D.}~\bibnamefont{Bryndin}},
  \bibinfo{author}{\bibfnamefont{E.~S.} \bibnamefont{Bozin}},
  \bibinfo{author}{\bibfnamefont{J.}~\bibnamefont{Bloch}},
  \bibinfo{author}{\bibfnamefont{T.}~\bibnamefont{Proffen}}, \bibnamefont{and}
  \bibinfo{author}{\bibfnamefont{S.~J.~L.} \bibnamefont{Billinge}},
  \bibinfo{journal}{J. Phys.: Condens. Matter} \textbf{\bibinfo{volume}{19}},
  \bibinfo{pages}{335219} (\bibinfo{year}{2007}).

\bibitem[{\citenamefont{Klemme and Miltenburg}(2004)}]{klemme_2004}
\bibinfo{author}{\bibfnamefont{S.}~\bibnamefont{Klemme}} \bibnamefont{and}
  \bibinfo{author}{\bibfnamefont{J.~C.~V.} \bibnamefont{Miltenburg}},
  \bibinfo{journal}{Mineral. Mag.} \textbf{\bibinfo{volume}{68}},
  \bibinfo{pages}{515} (\bibinfo{year}{2004}).

\bibitem[{\citenamefont{Lee et~al.}(2007)\citenamefont{Lee, Gasparovic,
  Broholm, Matsuda, Chung, Kim, Ueda, Xu, Zschack, Kakurai et~al.}}]{lee_2007}
\bibinfo{author}{\bibfnamefont{S.-H.} \bibnamefont{Lee}},
  \bibinfo{author}{\bibfnamefont{G.}~\bibnamefont{Gasparovic}},
  \bibinfo{author}{\bibfnamefont{C.}~\bibnamefont{Broholm}},
  \bibinfo{author}{\bibfnamefont{M.}~\bibnamefont{Matsuda}},
  \bibinfo{author}{\bibfnamefont{J.-H.} \bibnamefont{Chung}},
  \bibinfo{author}{\bibfnamefont{Y.~J.} \bibnamefont{Kim}},
  \bibinfo{author}{\bibfnamefont{H.}~\bibnamefont{Ueda}},
  \bibinfo{author}{\bibfnamefont{G.}~\bibnamefont{Xu}},
  \bibinfo{author}{\bibfnamefont{P.}~\bibnamefont{Zschack}},
  \bibinfo{author}{\bibfnamefont{K.}~\bibnamefont{Kakurai}},
  \bibnamefont{et~al.}, \bibinfo{journal}{J. Phys.: Condens. Matter}
  \textbf{\bibinfo{volume}{19}}, \bibinfo{pages}{145259}
  (\bibinfo{year}{2007}).

\bibitem[{\citenamefont{Menyuk et~al.}(1964)\citenamefont{Menyuk, Dwight, and
  Wold}}]{Menyuk_1964}
\bibinfo{author}{\bibfnamefont{N.}~\bibnamefont{Menyuk}},
  \bibinfo{author}{\bibfnamefont{K.}~\bibnamefont{Dwight}}, \bibnamefont{and}
  \bibinfo{author}{\bibfnamefont{A.}~\bibnamefont{Wold}}, \bibinfo{journal}{J.
  Phys.-Paris} \textbf{\bibinfo{volume}{25}}, \bibinfo{pages}{528}
  (\bibinfo{year}{1964}).

\bibitem[{\citenamefont{Tomiyasu et~al.}(2004)\citenamefont{Tomiyasu, Fukunaga,
  and Suzuki}}]{Tomiyasu_2004}
\bibinfo{author}{\bibfnamefont{K.}~\bibnamefont{Tomiyasu}},
  \bibinfo{author}{\bibfnamefont{J.}~\bibnamefont{Fukunaga}}, \bibnamefont{and}
  \bibinfo{author}{\bibfnamefont{H.}~\bibnamefont{Suzuki}},
  \bibinfo{journal}{Phys. Rev. B} \textbf{\bibinfo{volume}{70}},
  \bibinfo{pages}{214434} (\bibinfo{year}{2004}).

\bibitem[{\citenamefont{Lawes et~al.}(2006)\citenamefont{Lawes, Melot, Page,
  Ederer, Hayward, Proffen, and Seshadri}}]{lawes_2006}
\bibinfo{author}{\bibfnamefont{G.}~\bibnamefont{Lawes}},
  \bibinfo{author}{\bibfnamefont{B.~C.} \bibnamefont{Melot}},
  \bibinfo{author}{\bibfnamefont{K.}~\bibnamefont{Page}},
  \bibinfo{author}{\bibfnamefont{C.}~\bibnamefont{Ederer}},
  \bibinfo{author}{\bibfnamefont{M.~A.} \bibnamefont{Hayward}},
  \bibinfo{author}{\bibfnamefont{T.}~\bibnamefont{Proffen}}, \bibnamefont{and}
  \bibinfo{author}{\bibfnamefont{R.}~\bibnamefont{Seshadri}},
  \bibinfo{journal}{Phys. Rev. B} \textbf{\bibinfo{volume}{74}},
  \bibinfo{pages}{024413} (\bibinfo{year}{2006}).

\bibitem[{\citenamefont{Chang et~al.}(2009)\citenamefont{Chang, Huang, Li,
  Cheong, Ratcliff, and Lynn}}]{chang_2009}
\bibinfo{author}{\bibfnamefont{L.~J.} \bibnamefont{Chang}},
  \bibinfo{author}{\bibfnamefont{D.~J.} \bibnamefont{Huang}},
  \bibinfo{author}{\bibfnamefont{W.-H.} \bibnamefont{Li}},
  \bibinfo{author}{\bibfnamefont{S.-W.} \bibnamefont{Cheong}},
  \bibinfo{author}{\bibfnamefont{W.}~\bibnamefont{Ratcliff}}, \bibnamefont{and}
  \bibinfo{author}{\bibfnamefont{J.~W.} \bibnamefont{Lynn}},
  \bibinfo{journal}{J. Phys.: Condens. Matter} \textbf{\bibinfo{volume}{21}},
  \bibinfo{pages}{456008} (\bibinfo{year}{2009}).

\bibitem[{\citenamefont{Tsurkan et~al.}(2013)\citenamefont{Tsurkan, Zherlitsyn,
  Yasin, Felea, Skourski, Deisenhofer, von Nidda, Wosnitza, and
  Loidl}}]{tsurkan_2013}
\bibinfo{author}{\bibfnamefont{V.}~\bibnamefont{Tsurkan}},
  \bibinfo{author}{\bibfnamefont{S.}~\bibnamefont{Zherlitsyn}},
  \bibinfo{author}{\bibfnamefont{S.}~\bibnamefont{Yasin}},
  \bibinfo{author}{\bibfnamefont{V.}~\bibnamefont{Felea}},
  \bibinfo{author}{\bibfnamefont{Y.}~\bibnamefont{Skourski}},
  \bibinfo{author}{\bibfnamefont{J.}~\bibnamefont{Deisenhofer}},
  \bibinfo{author}{\bibfnamefont{H.-A.~K.} \bibnamefont{von Nidda}},
  \bibinfo{author}{\bibfnamefont{J.}~\bibnamefont{Wosnitza}}, \bibnamefont{and}
  \bibinfo{author}{\bibfnamefont{A.}~\bibnamefont{Loidl}},
  \bibinfo{journal}{Phys. Rev. Lett.} \textbf{\bibinfo{volume}{110}},
  \bibinfo{pages}{115502} (\bibinfo{year}{2013}).

\bibitem[{\citenamefont{Mufti et~al.}(2010)\citenamefont{Mufti, Nugroho, Blake,
  and Palstra}}]{mufti_2010}
\bibinfo{author}{\bibfnamefont{N.}~\bibnamefont{Mufti}},
  \bibinfo{author}{\bibfnamefont{A.~A.} \bibnamefont{Nugroho}},
  \bibinfo{author}{\bibfnamefont{G.~R.} \bibnamefont{Blake}}, \bibnamefont{and}
  \bibinfo{author}{\bibfnamefont{T.~T.~M.} \bibnamefont{Palstra}},
  \bibinfo{journal}{J. Phys.: Condens. Matter} \textbf{\bibinfo{volume}{22}},
  \bibinfo{pages}{075902} (\bibinfo{year}{2010}).

\bibitem[{\citenamefont{Ye et~al.}(1994)\citenamefont{Ye, Crottaz, Vaudano,
  Kubel, Tissot, and Schmid}}]{ye_1994}
\bibinfo{author}{\bibfnamefont{Z.-G.} \bibnamefont{Ye}},
  \bibinfo{author}{\bibfnamefont{O.}~\bibnamefont{Crottaz}},
  \bibinfo{author}{\bibfnamefont{F.}~\bibnamefont{Vaudano}},
  \bibinfo{author}{\bibfnamefont{F.}~\bibnamefont{Kubel}},
  \bibinfo{author}{\bibfnamefont{P.}~\bibnamefont{Tissot}}, \bibnamefont{and}
  \bibinfo{author}{\bibfnamefont{H.}~\bibnamefont{Schmid}},
  \bibinfo{journal}{Ferroelectr. Lett.} \textbf{\bibinfo{volume}{162}},
  \bibinfo{pages}{103} (\bibinfo{year}{1994}).

\bibitem[{\citenamefont{Prince}(1957)}]{prince_1957}
\bibinfo{author}{\bibfnamefont{E.}~\bibnamefont{Prince}},
  \bibinfo{journal}{Acta Crystallogr., Sect. C: Cryst. Struct. Commun.}
  \textbf{\bibinfo{volume}{10}}, \bibinfo{pages}{554} (\bibinfo{year}{1957}).

\bibitem[{\citenamefont{Rodriguez-Carvajal
  et~al.}(1998)\citenamefont{Rodriguez-Carvajal, Rousse, Masquelier, , and
  Hervieu}}]{rodriguez_1998}
\bibinfo{author}{\bibfnamefont{J.}~\bibnamefont{Rodriguez-Carvajal}},
  \bibinfo{author}{\bibfnamefont{G.}~\bibnamefont{Rousse}},
  \bibinfo{author}{\bibfnamefont{C.}~\bibnamefont{Masquelier}}, ,
  \bibnamefont{and} \bibinfo{author}{\bibfnamefont{M.}~\bibnamefont{Hervieu}},
  \bibinfo{journal}{Phys. Rev. Lett} \textbf{\bibinfo{volume}{81}},
  \bibinfo{pages}{4660} (\bibinfo{year}{1998}).

\bibitem[{\citenamefont{Melot et~al.}(2009{\natexlab{b}})\citenamefont{Melot,
  Page, Seshadri, Stoudenmire, Balents, Bergman, and Proffen}}]{Melot_2009b}
\bibinfo{author}{\bibfnamefont{B.~C.} \bibnamefont{Melot}},
  \bibinfo{author}{\bibfnamefont{K.}~\bibnamefont{Page}},
  \bibinfo{author}{\bibfnamefont{R.}~\bibnamefont{Seshadri}},
  \bibinfo{author}{\bibfnamefont{E.~M.} \bibnamefont{Stoudenmire}},
  \bibinfo{author}{\bibfnamefont{L.}~\bibnamefont{Balents}},
  \bibinfo{author}{\bibfnamefont{D.~L.} \bibnamefont{Bergman}},
  \bibnamefont{and} \bibinfo{author}{\bibfnamefont{T.}~\bibnamefont{Proffen}},
  \bibinfo{journal}{Phys. Rev. B} \textbf{\bibinfo{volume}{80}},
  \bibinfo{pages}{104420} (\bibinfo{year}{2009}{\natexlab{b}}).

\end{thebibliography}
\end{document}